\documentclass[10pt, fleqn]{article}

\usepackage{main_style}
\usepackage{notations}

\title{Blind calibration for compressed sensing:\\
  State evolution and an online algorithm}

\author[1]{Marylou Gabrié\thanks{Corresponding author: \href{mailto:marylou.gabrie@ens.fr}{marylou.gabrie@ens.fr}} }
\author[2]{Jean Barbier}
\author[1]{Florent Krzakala}
\author[3]{Lenka Zdeborová}

\affil[1]{Laboratoire de Physique de l’Ecole normale supérieure, ENS, Université PSL, CNRS, Sorbonne Université, Université de Paris, F-75005 Paris, France}
\affil[2]{International Center for Theoretical Physics, Trieste, Italy}
\affil[3]{Institut de Physique Théorique, CEA, CNRS, Université Paris-Saclay }

\begin{document}
\maketitle

\begin{abstract}
    Compressed sensing allows the acquisition of compressible signals with a small number of measurements. In experimental settings, the sensing process corresponding to the hardware implementation is not always perfectly known and may require a calibration. 
    To this end, blind calibration proposes to perform at the same time the calibration {\it and} the compressed sensing. Sch\"ulke and collaborators suggested an approach based on approximate message passing for blind calibration (cal-AMP) in \cite{Schulke2013,Schulke2015}. Here, their algorithm is extended from the already proposed offline case to the online case, for which the calibration is refined step by step as new measured samples are received. We show that the performance of both the offline and the online algorithms can be theoretically studied via the State Evolution (SE) formalism. 
    Finally, the efficiency of cal-AMP and the consistency of the theoretical predictions are confirmed through numerical simulations.
    
  \end{abstract}
  %

\section{Introduction}
The efficient acquisition of sparse signals has been made possible by Compressed Sensing (CS) \cite{Candes2006}. 
This technique has now many applications: in medical imaging \cite{Lustig2007,Otazo2010} for instance, where short acquisition times are preferable, or in imaging devices where measurements are costly \cite{Romberg2008}. For practical applications, correct calibration of potential unknowns coming from the measurement hardware is a central issue. If complementary pairs of measurement-signal are available a priori, a \emph{supervised} procedure of calibration can be imagined. We assume it is not the case and focus on \emph{blind} calibration. In blind calibration, calibration is conducted in parallel to the sparse signal recovery of the compressed sensing. This method does not require the availability of calibration samples before the reconstruction and is thus particularly appealing.
Here we consider formally the case where the exact measurement process is known up to a set of variables, referred to as calibration variables.

Gribonval and co-authors discussed in a series of articles \cite{Gribonval2012, Shen2013} a decalibration example where unknown gains multiply each sensor. More precisely, they considered for each signal $\x\in \R^N$ an observation $\y \in \R^M$, produced as
\begin{gather}\label{GLMmodel}
    \y = \mat{S} \, \W \, \x 
\end{gather}
with $\W \in \R^{M\times N}$ a known measurement matrix and $\mat{S}$ an unknown square diagonal matrix of calibration variables. The examined problem is to reconstruct both the corresponding signals $\x$ and the calibration matrix $\mat{S}$, provided several observations $\y$.
In \cite{Gribonval2012, Shen2013}, the authors showed that this question can be exactly expressed as a convex optimization problem, and thus be solved using off-the-shelf algorithms. In subsequent papers \cite{Schulke2013, Schulke2015}, Sch\"ulke and collaborators used instead an Approximate Message Passing (AMP) approach, famously introduced in compressed sensing by \cite{Donoho2009}. Their calibration-AMP algorithm (cal-AMP) was proposed for Bayesian blind calibration, with both the calibration variables on the sensors and the elements of the signals reconstructed simultaneously. Remarkably, cal-AMP is not restricted to the case where the distortion of the measurements is a multiplication by a gain and is applicable in other settings. 
The derivation of the algorithm relies on the extension of AMP resolution technics to the Generalized Linear Model (GLM), for which linear mixing is followed by a generic probabilistic sensing process, corresponding to the Generalized-AMP (GAMP) algorithm \cite{Kabashima2004, Rangan2011}.
Authors of \cite{Schulke2013,Schulke2015} demonstrated their approach through empirical numerical simulations, and found that considering relatively few measured samples at a time could already allow for a good calibration.

An ensuing question is whether the calibration could be performed \emph{online}, that is when different observations are received successively instead of being treated at once.
In learning applications, it is sometimes advantageous for speed concerns to only treat a subset of training examples at a time. Sometimes also, the size of the current data sets may exceed the available memory. Methods implementing a step-by-step learning, as the data arrives, are referred to as \emph{online}, \emph{streaming} or \emph{mini-batch} learning, as opposed to \emph{offline} or \emph{batch} learning.
For instance, in deep learning, Stochastic Gradient Descent is the most popular training algorithm \cite{Bottou2018}. From the theoretical point of view, the restriction to the fully online case, where a single data point is used at a time, offers interesting possibilities of analysis, as demonstrated in different machine learning problems by \cite{Wang2016,Mitliagkas2013,Wang2017}. 
Here we will consider the Bayesian online learning of the calibration variables. 


In the present paper, we revisit and extend cal-AMP with the following contributions
\begin{itemize}
    \item We re-derive the message passing equations from a more general formulation than \cite{Schulke2013,Schulke2015}. This strategy allows us to theoretically analyze the algorithm through a State Evolution, as first done for regular AMP in \cite{Donoho2009}. This analysis was not straightforward and remained an open problem for \cite{Schulke2015}. These contributions are presented in \citesec~\ref{sec:chap6-offline}.
    \item  In \citesec~\ref{sec:offline-fnrg}, we write the free energy, or equivalently the mutual information, of the problem thanks to the general formulation and the natural connection to committee machines \cite{Aubin2018}. We conjecture our results to be rigorously exact and discuss the missing ingredients to turn our formula into a full theorem.
    \item Along the lines of \cite{Manoel2018},
    we also consider the online version of the problem in \citesec~\ref{sec:chap6-online}. We propose an online cal-AMP algorithm allowing for the Bayesian adjustment of the calibration as new observations arrive. The corresponding State Evolution analysis is presented as well.
    \item  We validate numerically the performance of the algorithms and their consistency with the theoretical analyses, on the example of gain calibration, in \citesec~\ref{sec:chap6-numerics}. In particular, the algorithm remains fast and remarkably efficient going from the offline to the online setting. 
\end{itemize}

Finally, we wish to clarify notations: vector variables are underlined as $\x$, matrix variables are doubly underlined as $\W$ and the symbol $\propto$ omits the normalization factor of probability distributions. 
\section{Cal-AMP revisited}
\label{sec:chap6-offline}

\subsection{Model of calibrated generalized linear estimation}
\label{sec:chap6-model}
\begin{figure}[t]
    \centering
    \subfloat[]{\includegraphics[width=0.35\textwidth]{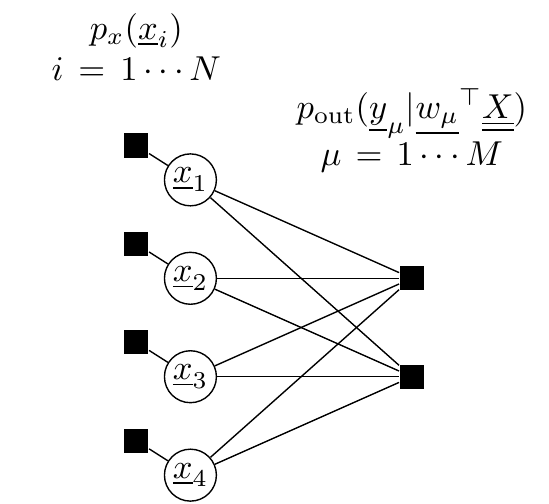}
    \label{fig:chap6-vect-amp}}
    \subfloat[.]{\includegraphics[width=0.4\textwidth]{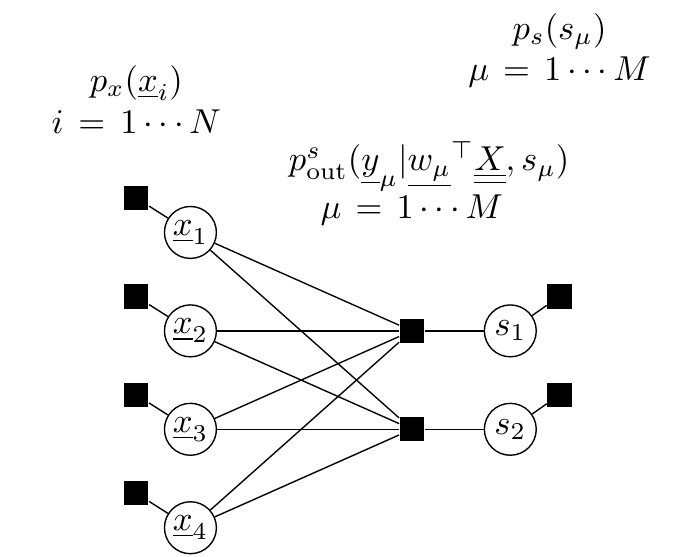}
    \label{fig:chap6-vect-amp-cal}}
    \caption{\textbf{(a)} Factor graph of the Generalized Linear Model (GLM) on vector variables corresponding to the joint distribution \eqref{eq:chap6-glm-vec-meas}. \textbf{(b)} Factor graph for the GLM on vector variables with not perfectly known channel including calibration variables, corresponding to the joint distribution \eqref{eq:chap6-glm-cal-meas}.}
\end{figure}

We consider $P$ observations $\y\kk \in \R^M$ gathered in a matrix $\Y \in \R^{M\times P}$, generated by the following \emph{teacher} model
\begin{align}
\hspace{-1cm}\vect{s}_0 &\sim p_{s_0}(\vect{s}_0) \\
    \forall \ k  = 1, \cdots,  P, \quad 
    \x_{0,(k)} &\sim p_{x_0}(\x_{0,(k)}) = \prod\limits_{i=1}^N p_{x_0}(x_{0,i, (k)}) \\
     \y\kk &\sim \poutos(\y\kk | \W\x_{0,(k)}, \vect{s}_0) = \prod\limits_{\mu = 1}^M \poutos(y_{(k),\mu}| \W\x_{0,(k)}, s_{0,\mu}) \,.
\end{align}
The $\x_{0,(k)} \in \R^N$ are $P$ unknown signals, gathered in a matrix $\X_0 \in \R^{N\times P}$, linearly mixed by a known weight (or measurement) matrix $\W$, and pushed through a noisy channel denoted $\poutos$.
The channel is not perfectly known and depends on the realization of calibration variables $\vect{s}_0 \in \R^M$ (which, in the special case of a linear model \eqref{GLMmodel}, is the diagonal of $\mat{S}$). 
We are interested in the estimation of the unknown signals and calibration variables. The priors and the channel of the teacher model are not necessarily available and we assume the following, a priori different, \emph{student} model,
\begin{align}
    p(\X, \vect{s} | \Y, \W) &= \frac{1}{\cZ(\Y, \W)} p_x(\X)p_s(\vect{s})\pouts(\Y | \W\X, \vect{s}) \\
    \label{eq:chap6-glm-cal-meas}
    & = \frac{1}{\cZ(\Y, \W)} \prod\limits_{i=1}^N p_x(\x_i)\prod_{\mu=1}^M p_s(s_\mu) \pouts(\y_\mu | \vect{w}_\mu\T\X, s_\mu), 
\end{align}
with $\x_i \in \R^P$, and $\y_\mu \in \R^P$. The corresponding factor graph is drawn on \citefig~\ref{fig:chap6-vect-amp-cal}.
Note that the distribution could be further factorized over the index $k$ of the $P$ observations. The corresponding message passing was derived in \cite{Schulke2013, Schulke2015}. Here we adopt instead the level of factorization above and derive the AMP algorithm on vector variables in $\R^P$. Thus cal-AMP can be seen as a special case of GAMP on vector variables as we will see in the next section. This point of view is key to obtain the State Evolution analysis of the calibration problem.

\subsection{Cal-AMP as a special case of GAMP}
In our derivation calibration-AMP (cal-AMP) is interpreted as a special case of Generalized Approximate Message Passing (GAMP) \cite{Rangan2011},  where (a) the algorithm acts simultaneously on the vector variables and (b) the channel includes a marginalization over the calibration variables.  For this first variation, the AMP algorithm on vector variables was already derived by \cite{Aubin2018}, to treat committee machines (two-layers neural networks \cite{Watkin1993}) without calibration. 
We shall recall here the main steps of the derivation, see \cite{Gabrie2019a} for a recent methodological introductory review to message passing algorithms and mean-field approximations in learning problems.

\subsubsection{AMP for reconstruction of multiple samples}
\label{sec:chap6-vect-amp}
The systematic procedure to write AMP for a given joint probability distribution consists in first writing BP on the factor graph, second project the messages on a parametrized family of functions to obtain the corresponding relaxed-BP and third close the equations on a reduced set of parameters by keeping only leading terms in the thermodynamic limit.

For the generic GLM, without the calibration variable, the posterior measure we are interested in is
\begin{gather}
    \label{eq:chap6-glm-vec-meas}
    p(\X | \Y, \W) = \frac{1}{\cZ(\Y, \W)} \prod_{i=1}^N p(\x_i)\prod_{\mu=1}^M \pout(\y_\mu | \vect{w}_\mu\T\X / \sqrt{N}), \quad \x_i \in \R^P, \quad \y_\mu \in \R^P,
\end{gather}
where the known entries of matrix $\W$ are drawn i.i.d from a standard normal distribution (the scaling in $1/\sqrt{N}$ is from now on made explicit). 
The corresponding factor graph is given on \citefig~\ref{fig:chap6-vect-amp}. 
We are considering the simultaneous reconstruction of $P$ signals $\x_{0,(k)} \in \R^N$ and therefore write the message passing on the variables $\x_i \in \R^P$.
The major difference with the fully factorized version on scalar variables of \cite{Schulke2013,Schulke2015} is that we will consider covariance matrices between variables coming from the $P$ observations instead of assuming complete independence. 

\paragraph{Belief propagation (BP)} 
We start with BP on the factor graph of \citefig~\ref{fig:chap6-vect-amp}. For all pairs of indices $i-\mu$, we define the update equations of messages 
\begin{align}
    \label{eq:chap6-bp-calamp-1}
    \msg{\tilde{m}^{(t)}}{\mu}{i} (\x_i) &= \frac{1}{\msg{\cZ}{\mu}{i}}
    \int \prod_{i'\neq i} \dd{\x_{i'}}\pout \left(\y_\mu |  \sum_j \frac{W_{\mu j}}{\sqrt{N}}\x_j\right) \prod_{i'\neq i} \msg{m^{(t)}}{i'}{\mu}(\x_{i'})\\
    \label{eq:chap6-bp-calamp-2}
    \msg{m^{(t+1)}}{i}{\mu} (\x_i) &= \frac{1}{\msg{\cZ}{i}{\mu}}
    p_x(\x_i)\prod_{\mu' \neq \mu} \msg{\tilde{m}^{(t)}}{\mu'}{i}(\x_i),
\end{align}
where $\msg{\cZ}{\mu}{i}$ and $\msg{\cZ}{i}{\mu}$ are normalizations that allow to interpret messages as probability distributions.
To improve readability, we drop the time indices in the following derivation, and only recover them in the final algorithm.

\paragraph{Relaxed BP (r-BP)} The second step of the derivation consists in developing messages up to order $O(1/N)$ as we take the thermodynamic limit $N \to + \infty$ (at fixed $\alpha = M/N$). At this order, we will find that it is consistent to consider the messages to be approximately Gaussian, i.e. characterized by their means and covariances. Thus we define
\begin{align}
\label{eq:chap6-vect-amp-rbp-def-xh}
\msg{\xh}{i}{\mu} &= \int \dd{\x} \x \; \msg{m}{i}{\mu}(\x) \\
\label{eq:chap6-vect-amp-rbp-def-Cx}
\msg{\Cx}{i}{\mu} &= \int \dd{\x}  \x \x^T \; \msg{m}{i}{\mu}(\x)
\end{align}
and
\begin{align}
\label{eq:chap6-vect-amp-rbp-def-w}
\displaystyle
\msg{\w}{\mu}{i} &= \sum_{i' \neq i} \frac{\Wuip}{\sqrt{N}}\msg{\xh}{i'}{\mu} \\
\label{eq:chap6-vect-amp-rbp-def-V}
\displaystyle
\msg{\V}{\mu}{i} &= \sum_{i' \neq i}\frac{\Wuip^2}{N} \msg{\Cx}{i'}{\mu}, 
\end{align}
where $\msg{\w}{\mu}{i}$ and $\msg{\V}{\mu}{i}$ are related to the intermediate variable $\z_\mu = \vect{w}_\mu\T \X$.

\subparagraph{Expansion of $\msgt{m}{\mu}{i}$ -} 
We define the Fourier transform $\hat{p}_{\rm out}$ of $\pout(\y_\mu|\z_\mu)$ with respect to its argument $\z_\mu = \vect{w}_\mu\T\X$,
\begin{gather}
    \hat{p}_{\rm out}(\y_\mu|\vect{\xi}_\mu) = \int \dd{\z_\mu} p_{\rm out}(\y_\mu | \z_\mu) \, e^{- i \vect{\xi}_\mu\T \z_\mu}.
\end{gather}
Using reciprocally the Fourier representation of $\pout(\y_\mu|\z_\mu)$,
\begin{gather}
    \pout(\y_\mu|\z_\mu) = \frac{1}{(2\pi)^M } \int \dd{\vect{\xi}_\mu} \hat{p}_{\rm out}(\y_\mu | \vect{\xi}_\mu) \, e^{i \vect{\xi}_\mu\T \z_\mu},
\end{gather}
we decouple the integrals over the different $\x_{i'}$ in \eqref{eq:chap6-bp-calamp-1},
\begin{align}
    \label{eq:chap6-deric-calamp-1}
    \msgt{m}{\mu}{i} (\x_i) &
    \propto  \int \dd{\vect{\xi}_\mu} \hat{p}_{\rm out} \left(\y_\mu | \vect{\xi}_\mu\right) e^{i \frac{\Wui}{\sqrt{N}}\vect{\xi}_\mu\T\x_i} \prod_{i'\neq i} \int  \dd{\x_{i'}} \msg{m}{i'}{\mu}(\x_{i'})e^{i \frac{\Wuip}{\sqrt{N}} \vect{\xi}_\mu\T\x_{i'}} \\
    \label{eq:chap6-deric-calamp-2}
    & \propto \int \dd{\vect{\xi}_\mu} \hat{p}_{\rm out} \left(\y_\mu | \vect{\xi}_\mu\right) e^{i \underline{\xi}\T \left(\frac{\Wui}{\sqrt{N}}\x_i + \msg{\w}{\mu}{i}\right) - \frac 1 2 \underline{\xi}\T \msg{{\V}^{-1}}{\mu}{i} \underline{\xi}},
\end{align}
where developing the exponentials of the product in \eqref{eq:chap6-bp-calamp-1}  allows to express the integrals over the $\x_{i'}$ as a function of the definitions \eqref{eq:chap6-vect-amp-rbp-def-w}-\eqref{eq:chap6-vect-amp-rbp-def-V}, before re-exponentiating to obtain the final result \eqref{eq:chap6-deric-calamp-2}.
Now reversing the Fourier transform and performing the integral over $\vect{\xi}$, we can further rewrite 
\begin{align}
    \msgt{m}{\mu}{i} (\x_i) &
    \propto  \int \dd{\z_\mu} \pout \left(\y_\mu | \z_\mu\right) e^{- \frac{1}{2} 
    \left( \z_\mu - \frac{\Wui}{\sqrt{N}}\x_i - \msg{\w}{\mu}{i}\right)\T
    \msg{{\V}^{-1}}{\mu}{i}
    \left( \z_\mu - \frac{\Wui}{\sqrt{N}}\x_i - \msg{\w}{\mu}{i}\right)
    } \\
    \label{eq:chap6-vect-amp-rbp-1}
    &
    \propto  \int \dd{\z_\mu} \mathbb{P}_{\rm out}(\z_\mu; \msg{\w}{\mu}{i}, \msg{\V}{\mu}{i} ) e^{ \left( \z_\mu - \msg{\w}{\mu}{i} \right)\T
    \msg{{\V}^{-1}}{\mu}{i}
    \frac{\Wui}{\sqrt{N}}\x_i 
    - \frac{\Wui^2}{2N}\x_i\T \msg{{\V}^{-1}}{\mu}{i} \x_i
    },
\end{align}
where we are led to introduce the \emph{output update functions},
\begin{align}
    \label{eq:chap6-vect-amp-Pout}
    \mathbb{P}_{\rm out}(\z_\mu; \msg{\w}{\mu}{i}, \msg{\V}{\mu}{i} ) &=  \pout \left(\y_\mu | \z_\mu\right) \cN(\z_\mu; \msg{\w}{\mu}{i}, \msg{\V}{\mu}{i} ) \, ,\\
    \label{eq:chap6-vect-amp-Zout}
    \Zout(\y_\mu , \msg{\w}{\mu}{i}, \msg{\V}{\mu}{i} ) &= \int \dd{\z_\mu} \pout \left(\y_\mu | \z_\mu\right) \cN(\z_\mu; \msg{\w}{\mu}{i}, \msg{\V}{\mu}{i} ) \, ,\\
    \label{eq:chap6-vect-amp-gout-dgout}
    \gout(\y_\mu , \msg{\w}{\mu}{i}, \msg{\V}{\mu}{i} ) &= \frac{1}{\Zout} \frac{\partial \Zout}{\partial \w} \quad \text{ and } \quad
    \dgout = \frac{\partial \gout}{\partial \w},
\end{align}
with $\cN(\z;\w,\V)$ the multivariate Gaussian distribution of mean $\w$ and covariance $\V$.
Further expansion of the exponential in \eqref{eq:chap6-vect-amp-rbp-1} to order $O(1/N)$ leads to the Gaussian parametrization 
\begin{align}
    \msgt{m}{\mu}{i} (\x_i) & \propto 1 + \frac{{\Wui}}{\sqrt{N}} \gout \x_i + \frac{{{\Wui}}^2}{2 N} {\x_i}^T (\gout\gout^T + \dgout) \x_i \\
    & \propto e^{{\msg{\B}{\mu}{i}}^T\x_i - \frac 1 2 {\x_i}^T\msg{\A}{\mu}{i}\x_i}, 
\end{align}
with
\begin{align}
    \label{eq:chap6-vect-amp-rb-B}
    \msg{\B}{\mu}{i}  &= \frac{{\Wui}}{\sqrt{N}} \gout (\y_\mu , \msg{\w}{\mu}{i}, \msg{\V}{\mu}{i} ) \\
    \label{eq:chap6-vect-amp-rb-A}
    \msg{\A}{\mu}{i}  &= - \frac{{{\Wui}}^2}{ N} \dgout(\y_\mu , \msg{\w}{\mu}{i}, \msg{\V}{\mu}{i} ) .
\end{align}

\subparagraph{Consistency with $\msg{m}{i}{\mu}$ -}
Inserting the Gaussian approximation of $\msgt{m}{\mu}{i}$ in the definition of $\msg{m}{i}{\mu}$, we get the parametrization
\begin{align}
    \msg{m}{i}{\mu}(\x_i) & \propto p_x(\x_i) \prod_{\mu' \neq \mu} e^{{\msg{\B}{\mu'}{i}}^T\x_i - \frac 1 2 {\x_i}^T\msg{\A}{\mu'}{i}\x_i}  \propto p_x(\x_i) e^{-\frac{1}{2}(\x_i - \msg{\lbd}{i}{\mu})^T \msg{\sig}{i}{\mu}^{-1} (\x_i - \msg{\lbd}{i}{\mu})}
\end{align}
with
\begin{align}
    \label{eq:chap6-vect-amp-rbp-lbd}
\msg{\lbd}{i}{\mu}  &= \msg{\sig}{i}{\mu}\left( \sum_{\mu' \neq \mu} \msg{\B}{\mu'}{i} \right)  \\
\label{eq:chap6-vect-amp-rbp-sig}
\msg{\sig}{i}{\mu}  &= \left( \sum_{\mu' \neq \mu} \msg{\A}{\mu'}{i} \right)^{-1} .
\end{align} 

\subparagraph{Closing the equations -}
Ensuring the consistency with the definitions \eqref{eq:chap6-vect-amp-rbp-def-xh}-\eqref{eq:chap6-vect-amp-rbp-def-Cx} of mean and covariance of $\msg{m}{i}{\mu}$ we finally close our set of equations by defining the \emph{input update functions},
\begin{align}
    \label{eq:chap6-vect-amp-Zx}
    \cZ^x &= \int \dd{\x} p_x(\x)e^{-\frac 1 2 (\x-\lbd)\T\sig^{-1}(\x-\lbd)} \\
    \label{eq:chap6-vect-amp-f1x}
    \vect{f}^x_1(\lbd, \sig) &= \frac{1}{\cZ^x}\int \dd{\x} \x \, p_x(\x)e^{-\frac 1 2 (\x-\lbd)\T\sig^{-1}(\x-\lbd)} \\
    \label{eq:chap6-vect-amp-f2x}
    \mat{f}^x_2(\lbd, \sig) &=  \frac{1}{\cZ^x} \int \dd{\x} \x\x\T \, p_x(\x)e^{-\frac 1 2 (\x-\lbd)\T\sig^{-1}(\x-\lbd)} - \vect{f}^x_1(\lbd, \sig)\vect{f}^x_1(\lbd, \sig)\T,
\end{align}
so that
\begin{align}
    \label{eq:chap6-vect-amp-rb-xh}
    \msg{\xh}{i}{\mu}  &=  \vect{f}^x_1(\msg{\lbd}{i}{\mu} , \msg{\sig}{i}{\mu}) \\
    \label{eq:chap6-vect-amp-rb-Cx}
    \msg{\Cx}{i}{\mu}  &= \mat{f}^x_2(\msg{\lbd}{i}{\mu} , \msg{\sig}{i}{\mu}) .
\end{align}

The closed set of equations \eqref{eq:chap6-vect-amp-rbp-def-w}, \eqref{eq:chap6-vect-amp-rbp-def-V},  \eqref{eq:chap6-vect-amp-rb-B} \eqref{eq:chap6-vect-amp-rb-A}, \eqref{eq:chap6-vect-amp-rbp-lbd}, \eqref{eq:chap6-vect-amp-rbp-sig}, \eqref{eq:chap6-vect-amp-rb-xh} and \eqref{eq:chap6-vect-amp-rb-Cx}, with restored time indices, defines the r-BP algorithm. At convergence of the iterations, we obtain the approximated marginals
\begin{align}
    \label{eq:chap6-vect-amp-marginal-def}
    m_i(\x_i) = \frac 1 {\cZ_i^x} p_x(\x_i) e^{-\frac 1 2 (\x-\lbd_i)\T\sig_i^{-1}(\x-\lbd_i)} 
\end{align}
with 
\begin{align}
    \lbd_i  &= \sig_i\left( \sum\limits_{\mu=1}^M \msg{\B}{\mu}{i} \right)  \\
    \sig_i  &= \left( \sum\limits_{\mu}^M \msg{\A}{\mu}{i} \right)^{-1} .
\end{align}

While BP requires to track the iterations of $M \times N$ messages (distributions) over vectors in $\R^P$, r-BP only requires to track $O(M \times N \times P)$ variables, which is a great simplification. Nonetheless, r-BP is scarcely used as such. The computational cost can be readily further reduced with little more approximation. In the thermodynamic limit, the messages are closely related to the marginals as the contribution of the missing message between \eqref{eq:chap6-bp-calamp-2} and \eqref{eq:chap6-vect-amp-marginal-def} is negligible to a certain extent. Careful book keeping of the order of contributions of the different terms leads to a set of closed equations on parameters of the marginals, i.e. $O(N)$ variables, corresponding to the GAMP algorithm.

\paragraph{Approximate message passing}
Given the scaling of the weights in $O(1/\sqrt{N})$, the r-BP algorithm simplifies in the thermodynamic limit.
We define the pairs of marginal parameters $\w_\mu$, $\V_\mu$ and $\xh_i$, $\Cx_i$ similarly to the parameters $\lbd_i$ and $\sig_i$ defined above. The expansion of the original message parameters $\msg{\lbd}{i}{\mu}$, $\msg{\sig}{i}{\mu}$, $\msg{\w}{\mu}{i}$, $\msg{\V}{\mu}{i}$, $\msg{\xh}{i}{\mu}$ and $\msg{\Cx}{i}{\mu}$ around these new parameters leads to the final closed set of equations corresponding to the approximate message passing. In fine, we obtained the vectorized AMP for the GLM presented in \citealg~\ref{alg:chap6-vect-amp}. 

Note that, similarly to GAMP, relaxing the Gaussian assumption on the weight matrix entries to any distribution with finite second moment yields the same algorithm using the Central Limit Theorem. Furthermore, the algorithmic procedure should also generalize to a wider class of random matrices allowing for correlations between entries: the ensemble of orthogonally invariant matrices. In the singular value decomposition of such weight matrices $\W=\mat{U}\,\mat{S}\,\mat{V}\T \in \R^{M\times N}$ the orthogonal basis matrices $\mat{U}$ and $\mat{V}$ are drawn uniformly at random from respectively $\mathrm{O}(M)$ and $\mathrm{O}(N)$, while the diagonal matrix of singular values $\mat{S}$ has an arbitrary spectrum. For the GLM, without calibration variables, the signal is recovered in such cases by the (Generalized) Vector-Approximate Message Passing (G-VAMP) algorithm \cite{Rangan2016,Schniter2016}, inspired from prior works in statistical physics \cite{Opper2001,Opper2001prl,Kabashima2008, Shinzato2009, Kabashima2014} and statistical inference \cite{Minka2001, Opper2005}.

\begin{algorithm*}[htb]
\caption{Generalized Approximate Message Passing  (GAMP) for vectors \label{alg:chap6-vect-amp}}   
\begin{algorithmic}
    \State {\bfseries Input:} matrix $\Y \in \R^{M  \times P}$ and matrix $\W \in \R^{M \times N}$:
    \State \emph{Initialize}: $\xh_i$, $\Cx_i, \sig_i \quad \forall i$ and $\gout_\mu$, $\dgout_\mu \quad \forall \mu$
    \Repeat   
    \Statex 1) Estimate mean and variance of $\z_\mu$ given current $\xh_i$
    \vspace{-0.3cm}
    \begin{align}
        \V_{\mu}^{(t)} &= \sum\limits_{i=1}^N \frac{W_{\mu i}^2}{N} {\Cx_{i}}^{(t)} \label{alg:chap6-vect-amp-V} \\
        \w_{\mu}^{(t)} &= \sum\limits_{i = 1}^N \frac{W_{\mu i}}{\sqrt{N}} \xh^{(t)}_i - \sum\limits_{i = 1}^N \frac{W_{\mu i}^2}{N} (\sig_i^{(t)})^{-1} {\Cx}^{(t)}_i \sig_i \gout_\mu^{(t-1)} \label{alg:chap6-vect-amp-om}
    \end{align} 
    \vspace{-0.3cm}
    \Statex 2) Estimate mean and variance of the gap between optimal $\z_\mu$ and $\w_\mu$ given $\y_\mu$
    \vspace{-0.3cm}
    \begin{align}
        \dgout^{(t)}_\mu &=  \dgout( \y_\mu, \w^{(t)}_\mu, \V^{(t)}_\mu) \label{alg:chap6-vect-amp-dg} \\
        \gout^{(t)}_\mu &= \gout( \y_\mu, \w^{(t)}_\mu, \V^{(t)}_\mu)  \label{alg:chap6-vect-amp-g}
    \end{align}
    \vspace{-0.3cm}
    \Statex 3) Estimate mean and variance of $\x_i$ given current optimal $\z_\mu$
    \vspace{-0.3cm}
    \begin{align}
        \sig_i^{(t)} &= \left(- \sum\limits_{\mu=1}^{M}\frac{W_{\mu i}^2}{N}\dgout_\mu^{(t)}\right)^{-1} \label{alg:chap6-vect-amp-sig} \\
        \lbd_i^{(t)} &=  \xh^{(t)}_i + \sig_i^{(t)}\left( \sum\limits_{\mu=1}^{M}\frac{W_{\mu i}}{\sqrt{N}}\gout_\mu^{(t)}\right) \label{alg:chap6-vect-amp-lbd}
    \end{align}
    \vspace{-0.3cm}
    \Statex 4) Estimate of mean and variance of  $\x_i$ augmented of the information about the prior
    \vspace{-0.3cm}
    \begin{align}
        {\Cx_i}^{(t+1)} &= \mat{f}^x_2(\lbd^{(t)}_i, \sig_i^{(t)}) \label{alg:chap6-vect-amp-cx} \\
        \xh^{(t+1)}_i &= \vect{f}^x_1(\lbd^{(t)}_i, \sig_i^{(t)})  \label{alg:chap6-vect-amp-xh}
    \end{align}

    \vspace{0.05cm}
    \Until{convergence} 
\end{algorithmic}
\end{algorithm*}

\subsubsection{Treatment of calibration variables}
\paragraph{Heuristic derivation}
To deal with calibration variables, we need to consider the factor graph of \citefig~\ref{fig:chap6-vect-amp-cal}. Here belief propagation must involve a set of messages related to the variables $s_\mu$. This new algorithm can easily be deduced from the GAMP algorithm derived above by noticing that 
the posterior distribution on $\X$ in the presence of the calibration variable $\vect{s}$ is a special case of the GLM on vector variables with the effective channel
\begin{gather}
    \pout(\y_\mu|\vect{w_\mu}\T\X) = \int \dd{s_\mu}\pouts(\y_\mu|\vect{w_\mu}\T\X, s_\mu)p_s(s_\mu).
\end{gather}
Thus \citealg~\ref{alg:chap6-vect-amp} can reconstruct $\X$ using output functions \eqref{eq:chap6-vect-amp-Pout}-\eqref{eq:chap6-vect-amp-gout-dgout} which include a marginalization over $\vect{s}$:
\begin{gather}
    \Zout(\y_\mu , \w_\mu, \V_\mu ) = \int \dd{\z_\mu} \int \dd{s_\mu} \pouts \left(\y_\mu | \z_\mu, s_\mu \right) p_s(s_\mu)\cN(\z_\mu; \w_\mu, \V_\mu ) .
\end{gather}
The estimator of a variable computed by (sum-product) AMP is always the mean of an approximate posterior marginal distribution. For the calibration variable, the AMP posterior is already displayed in the above $\Zout$, 
\begin{gather}
    \label{eq:chap6-vect-amp-post-s}
    m^s_\mu(s_\mu) = \frac 1 {\Zout} \int \dd{\z_\mu} \pouts \left(\y_\mu | \z_\mu, s_\mu \right)p_s(s_\mu) \cN(\z_\mu; \w_\mu, \V_\mu ).
\end{gather}  
As a result, we can compute the estimate and uncertainty on the calibration variable, at convergence of \citealg~\ref{alg:chap6-vect-amp}, as
\begin{align}
    \label{eq:chap6-vect-amp-sh}
    \sh_\mu &= f_1^s(\y_\mu, \w_\mu, \V_\mu) = \frac 1 {\Zout} \int \dd{s_\mu} s_\mu \int \dd{\z_\mu} \pouts \left(\y_\mu | \z_\mu, s_\mu \right)p_s(s_\mu) \cN(\z_\mu; \w_\mu, \V_\mu ) \\
    \Cs_\mu&= f_2^s(\y_\mu, \w_\mu, \V_\mu) = \frac 1 {\Zout} \int \dd{s_\mu} s_\mu^2 \int \dd{\z_\mu} \pouts \left(\y_\mu | \z_\mu, s_\mu \right)p_s(s_\mu) \cN(\z_\mu; \w_\mu, \V_\mu ) - \sh_\mu^2 . 
\end{align}

\paragraph{Relation to original cal-AMP derivation}
In \cite{Schulke2013, Schulke2015}, the cal-AMP algorithm was derived from the belief propagation equations on the scalar variables of the fully factorized distribution (over $N$, $M$ and $P$). It is equivalent to \citealg~\ref{alg:chap6-vect-amp} if the covariance matrices $\V_\mu$, $\dgout_\mu$, $\sig_i$, $\Cx_i$ are assumed to be diagonal. 
However, we recall that BP is only exact on a tree, where the incoming messages at each node are truly independent. On the dense factor graph of the GLM on scalar variables, this is approximately exact in the thermodynamic limit due to the random mixing and small scaling of the weight matrix $\W$. Here, by considering the reconstruction of $P$ samples at the same time, sharing given realizations of the calibration variable $\vect{s}$, as well as the measurement matrix $\W$, additional correlations may arise. Writing the message passing on the vector variables in $\R^P$ allows not to neglect them and therefore improve the final algorithm.

\subsection{State Evolution for cal-AMP}
In the large size limit, where $N \to + \infty$ at fixed $\alpha = M/N$, the performance of the AMP algorithm can be characterized by a set of simpler equations corresponding to a quenched average over the disorder (here the realizations of $\X_0$, $s_0$, $\Y$ and $\W$), referred to as State Evolution (SE) \cite{Donoho2009}. 
Remarkably, the SE equations are equivalent to the saddle point equations of the replica free energy associated with the problem under the Replica Symmetric (RS) ansatz \cite{Zdeborova2016}.  
In \cite{Aubin2018}, the teacher-student matched setting of the GLM on vectors is examined through the replica approach and the Bayes optimal State Evolution equations are obtained through this second strategy. In the following we present the alternative derivation of the State Evolution equations from the message passing without assuming a priori matched teacher and student models. Our starting point are the r-BP equations. Finally, we also introduce new State Evolution equations following the reconstruction of calibration variables.

\subsubsection{State Evolution derivation for mismatched prior and channel}
\label{sec:chap6-se-cal-amp}
\paragraph{Definition of the overlaps}
The key quantities to track the dynamic of iterations and the fixed point(s) of AMP are the overlaps. Here, they are the $P \times P$ matrices
\begin{gather}
    \q = \frac{1}{N} \sum_{i=1}^N \xh_i \xh_i^T, \quad \mm = \frac{1}{N} \sum_{i=1}^N \xh_i {\x_{0,i}}^T, \quad \q_0 = \frac{1}{N} \sum_{i=1}^N \x_{0,i} {\x_{0,i}}^T.
\end{gather} 

\paragraph{Output parameters}
Under independent statistics of the entries of $\W$ and under the assumption of independent incoming messages, the variable $\msg{\w}{\mu}{i}$ defined in \eqref{eq:chap6-vect-amp-rbp-def-w} is a sum of independent variables and follows a Gaussian distribution by the Central Limit Theorem. Its first and second moments are
\begin{align}
    \EE{\W}{\msg{\w}{\mu}{i}} & = \frac{1}{\sqrt{N}} \sum_{i'\neq i} \EE{\W}{W_{\mu i}} \msg{\xh}{i'}{\mu} = 0 \, ,\\
\EE{\W}{\msg{\w}{\mu}{i} \msg{\w}{\mu}{i} ^T} 
	& =  \frac{1}{N} \sum_{i'\neq i} \EE{\W}{W^2_{\mu i}} \msg{\xh}{i'}{\mu}\msg{\xh}{i'}{\mu}^T \\
    &=  \frac{1}{N} \sum_{i=1}^N \msg{\xh}{i}{\mu}\msg{\xh}{i}{\mu}^T + O\left({1}/{N}\right)  \\
	& =  \frac{1}{N} \sum_{i=1}^N \xh_i\xh_i^T - \partial_{\lbd} \vect{f}^x_1\sig_i\msg{B}{\mu}{i} \xh_i^T - \left(\partial_{\lbd} \vect{f}_1^x\sig_i\msg{B}{\mu}{i} \xh_i^T \right)^T +O\left({1}/{N}\right) \\
	& = \frac{1}{N} \sum_{i=1}^N \xh_i\xh_i^T + O\left({1}/{\sqrt{N}}\right)\,
\end{align}
where we used the fact that $\msg{B}{\mu}{i}$ defined in \eqref{eq:chap6-vect-amp-rb-B} is of order $O(1/\sqrt{N})$.
Similarly, the variable $\msg{\z}{\mu}{i} = \sum_{i'\neq i} \frac{W_{\mu i'}}{\sqrt{N}} \x_{i'}$ is Gaussian with first and second moments
\begin{align}
    \EE{\W}{\msg{\z}{\mu}{i}} 
    &= \frac{1}{\sqrt{N}} \sum_{i'\neq i} \EE{\W}{W_{\mu i}} \x_{0,i'} = 0 \, ,
    \\
    \EE{\W}{\msg{\z}{\mu}{i} \msg{\z}{\mu}{i} ^T} 
        &= \frac{1}{N} \sum_{i=1}^N \x_{0,i}{\x_{0,i}}^T + O\left({1}/{\sqrt{N}}\right). 
    \end{align}
Furthermore, the covariance between $\msg{\w}{\mu}{i}$ and $\msg{\z}{\mu}{i}$ is 
\begin{align}
    \EE{\W}{\msg{\z}{\mu}{i} \msg{\w}{\mu}{i}^T} 
        & =  \frac{1}{N} \sum_{i'\neq i} \EE{\W}{W^2_{\mu i}} \x_{0,i'}\msg{\xh}{i'}{\mu}^T \\
        &=  \frac{1}{N} \sum_{i=1}^N \x_{0,i'}\msg{\xh}{i}{\mu}^T + O\left({1}/{N}\right)  \\
        & =  \frac{1}{N} \sum_{i=1}^N \x_{0,i'}\xh_i^T -  \x_{0,i'}\partial_{\lbd} \vect{f}^x_1\sig_i\msg{\B}{\mu}{i}^T +O\left({1}/{N}\right) \\
        & = \frac{1}{N} \sum_{i=1}^N \x_{0,i'}\xh_i^T + O\left({1}/{\sqrt{N}}\right). 
    \end{align} 
Hence we find that for all $\mu$-s and all $i$-s, $  \msg{\w}{\mu}{i}$ and $ \msg{\z}{\mu}{i}$ are approximately jointly Gaussian in the thermodynamic limit following a unique distribution $\cN\left( \msg{\z}{\mu}{i}, \msg{\w}{\mu}{i}; \; \vect{0}, \, \Q \right) $ with the block covariance matrix
\begin{gather}
        \Q = 
        \begin{bmatrix}
        \q_0 & \mm \\
        \\
        {\mm}\T & \q \\ 
        \end{bmatrix}.
\end{gather}
For the variance message $\msg{\V}{\mu}{i}$, defined in \eqref{eq:chap6-vect-amp-rbp-def-V}, we have 
\begin{align}
    \EE{\W}{\msg{\V}{\mu}{i}} &= \sum_{i'\neq i} \EE{\W}{\frac{W_{\mu i}}{N}^2} \msg{\Cx}{i'}{\mu} \\
    &= \sum_{i=1}^N \frac{1}{N} \msg{\Cx}{i}{\mu} + O\left({1}/{N}\right) \\
    &= \sum_{i=1}^N \frac{1}{N} \Cx_i + O\left({1}/{\sqrt{N}}\right) ,
\end{align}
where using the developments of $\msg{\lbd}{i}{\mu}$ and $\msg{\sig}{i}{\mu}$ \eqref{eq:chap6-vect-amp-rbp-lbd}-\eqref{eq:chap6-vect-amp-rbp-sig} and the scaling of $\msg{\B}{\mu}{i} $ in $O({1}/{\sqrt{N}})$ we replaced 
\begin{align}
    \msg{\Cx}{i}{\mu} = \mat{f}_2^x(\msg{\lbd}{i}{\mu}, \msg{\sig}{i}{\mu}) = \mat{f}_2^x(\lbd_i, \sig_i) - \partial_{\lbd}\mat{f}^x_2 \sig_i \msg{\B}{\mu}{i}^T = \mat{f}_2^x(\lbd_i, \sig_i) + O\left({1}/{\sqrt{N}}\right).
\end{align}
Futhermore, we can check that under our assumptions
\begin{gather}
    \lim_{N\to + \infty} \EE{\W}{\msg{\V}{\mu}{i}^2 - \EE{\W}{\msg{\V}{\mu}{i}}^2} = 0 ,
\end{gather}
where the square is taken element-wise, meaning that all $\msg{\V}{\mu}{i}$ concentrate on their mean in the thermodynamic limit, which is identical for all of them:
\begin{gather}
    \V = \sum_{i=1}^N \frac{1}{N} \Cx_i .
\end{gather}

\paragraph{Input parameters} Here we use the re-parametrization trick to express $\y_\mu$ as a function $g_0(\cdot)$ taking the calibration variable $s_\mu$ and a noise $\eps_\mu \sim p_\epsilon(\eps_\mu)$ as inputs:
    $\y_\mu = g_0(\vect{w}_\mu\T\X_0, s_{0,\mu}, \eps_\mu)$.
Following \eqref{eq:chap6-vect-amp-rb-A}-\eqref{eq:chap6-vect-amp-rb-B} and \eqref{eq:chap6-vect-amp-marginal-def}, 
\begin{align}
    \sig_i^{-1}\lbd_i 
    & = \sum_{\mu=1}^M \frac{W_{\mu i}}{\sqrt{N}} \gout\left(\y_\mu, \msg{\w}{\mu}{i},  \msg{\V}{\mu}{i}\right) \\
    & =  \sum_{\mu=1}^M \frac{W_{\mu i}}{\sqrt{N}} \gout\left(g_0\left( \sum_{i'\neq i} \frac{W_{\mu i'}}{\sqrt{N}} \x_{0,i'} + \frac{W_{\mu i}}{\sqrt{N}} \x_{0,i},s_{0,\mu}, \eps_\mu \right), \msg{\w}{\mu}{i},  \msg{\V}{\mu}{i}\right) \\
    & =  \sum_{\mu=1}^M \frac{W_{\mu i}}{\sqrt{N}} \gout\left(g_0\left( \sum_{i'\neq i} \frac{W_{\mu i'}}{\sqrt{N}} \x_{0,i'}, s_{0,\mu}, \eps_\mu \right), \msg{\w}{\mu}{i},  \msg{\V}{\mu}{i}\right)  \notag\\
    &\qquad \qquad + \sum_{\mu=1}^M \frac{W^2_{\mu i}}{N}  \partial_{z} \mat{\gouts}\left(g_0\left( \msg{\z}{\mu}{i}, s_{0,\mu}, \eps_\mu\right), \msg{\w}{\mu}{i}, \msg{\V}{\mu}{i}\right) \x_{0,i}.
\end{align}
The first term is again a sum of independent random variables, given the $W_{\mu i}$ are i.i.d. with zero mean and the messages of type $\mu \to i$ are assumed independent. The second term has non-zero mean and can be shown to concentrate. Recalling that all $\msg{\V}{\mu}{i}$ also concentrate on $\V$, we obtain that $\sig_i^{-1}\lbd_i$ effectively follows the Gaussian distribution ($I_P$ is the $P\times P$ identity)
\begin{gather}
    \sig_i^{-1}\lbd_i \sim \cN\left(\sig_i^{-1}\lbd_i;\; \alpha \mh \x_{0,i}, \sqrt{\alpha \qh} I_P\right) 
\end{gather}
with
\begin{align}
    \label{eq:chap6-se-non-nishi-qh}
    \qh &= \int \dd{\eps} p_{\epsilon}(\eps) \dd{s_0}p_{s_0}(s_0) \int \dd{\w} \dd{\z} \cN(\z, \w ; \underline{0}, \Q) 
    \gout(g_0\left( \z  , s_0,  \eps\right) ,\w,  \V) \times \\
    &\qquad \qquad \qquad \qquad \qquad \qquad \qquad \qquad \qquad \qquad \qquad \qquad \qquad \gout(g_0\left( \z  , s_0, \eps\right), \w,  \V)^T \, ,\notag \\
	\label{eq:chap6-se-non-nishi-mh}
    \mh  &= \int \dd{\eps} p_{\epsilon}(\eps) \dd{s_0}p_{s_0}(s_0) \int \dd{\w} \dd{\z} \cN(\z, \w ; \underline{0}, \Q)
	\partial_{\z} \mat{\gouts}(g_0\left( \z  , s_0,  \eps\right), \w , \V) .
\end{align}
The inverse variance $\sig_i^{-1}$ can be shown to concentrate on its mean 
\begin{align}
    \sig_i^{-1}  &= \sum\limits_{\mu=1}^M \frac{{\Wui}^2}{ N} \dgout(\y_\mu , \msg{\w}{\mu}{i}, \msg{\V}{\mu}{i} ) \simeq \alpha \chih \, , \\
    \label{eq:chap6-se-non-nishi-chih}
    \chih  &= - \int \dd{\eps} p_\epsilon(\eps) \dd{s_0} p_{s_0}(s_0)\int \dd{\eps} \dd\z \cN(\z, \w ; \underline{0}, \Q) 
	 \partial_{\omega} \mat{\gouts}( g_0\left( \z , s_0, \eps\right), \w , \V) \, .
\end{align}

\paragraph{Closing the equations} The statistics of the input parameters $\lbd_i$ and $\sig_i$ must be consistent with the overlap matrices given
\begin{align}
    \V &= \frac{1}{N}\sum\limits_{i=1}^N \Cx_i = \EE{\lbd, \sig}{\mat{f}^x_2(\lbd, \sig)} ,\\
    \q &= \frac{1}{N} \sum\limits_{i=1}^N \xh_i \xh_i\T = \EE{\lbd, \sig}{\mat{f}^x_1(\lbd, \sig)\mat{f}^x_1(\lbd, \sig)\T},\\
    \mm &= \frac{1}{N} \sum\limits_{i=1}^N \xh_i {\x_{0,i}}\T = \EE{\lbd, \sig}{ \mat{f}^x_1(\lbd, \sig){\x_{0,i}}\T } ,
\end{align}
which gives upon expressing the computation of the expectations (here $ \D{\vect{\xi}}$ is the $P$-dimensional multivariate standard Gaussian distribution with $\underline{0}$ mean and identity covariance)
\begin{align}
    \label{eq:chap6-se-non-nishi-V}
    \V &= \int \dd{\x_0}p_{x_0}(\x_0) \int \D{\vect{\xi}} \mat{f}^x_2 \left( (\alpha \chih)^{-1}\left({\sqrt{\alpha \qh} \underline{\xi} + \alpha \mh\x_0}\right); (\alpha \chih)^{-1}  \right) \, , \\
    \label{eq:chap6-se-non-nishi-m}
    \mm &= \int \dd{\x_0}p_{x_0}(\x_0) \int \D{\vect{\xi}} \vect{f}^x_1 \left( (\alpha \chih)^{-1}\left({\sqrt{\alpha \qh} \underline{\xi} + \alpha \mh\x_0}\right); (\alpha \chih)^{-1}  \right){\x_0}\T \, , \\
    \q &=  \int \dd{\x_0}p_{x_0}(\x_0) \int \D{\vect{\xi}} \vect{f}^x_1 \left( (\alpha \chih)^{-1}\left({\sqrt{\alpha \qh} \underline{\xi} + \alpha \mh\x_0}\right); (\alpha \chih)^{-1}  \right) \times \notag\\
    \label{eq:chap6-se-non-nishi-q}
    &\qquad \qquad \qquad \qquad \qquad \qquad \qquad \qquad \vect{f}^x_1 \left( (\alpha \chih)^{-1}\left({\sqrt{\alpha \qh} \underline{\xi} + \alpha \mh\x_0}\right); (\alpha \chih)^{-1}  \right)\T .
\end{align}
The State Evolution analysis of the GLM with vector variables, that allows to track the vectorial version of GAMP given by Algorithm~\ref{alg:chap6-vect-amp}, finally consists in iterating alternatively the equations \eqref{eq:chap6-se-non-nishi-qh}, \eqref{eq:chap6-se-non-nishi-mh}, \eqref{eq:chap6-se-non-nishi-chih}, and the equations \eqref{eq:chap6-se-non-nishi-V}, \eqref{eq:chap6-se-non-nishi-m} \eqref{eq:chap6-se-non-nishi-q} until convergence.

\paragraph{Reconstruction of the calibration variable} 
In parallel, the reconstruction of $\vect{s}_0$ can be followed by introducing the scalar overlaps
\begin{gather}
    r = \frac{1}{M} \sum_{\mu=1}^M \sh_\mu ^2 , \quad \nu = \frac{1}{M} \sum_{\mu=1}^M \sh_\mu s_{0,\mu},  \quad r_0 = \frac{1}{M} \sum_{\mu=1}^M s^2_{0,\mu} .
\end{gather}
Recalling the definition of the estimator $\vect{\sh}(\cdot)=\vect{f}_1^s(\cdot)$ given by \eqref{eq:chap6-vect-amp-sh} and taking the steps of the derivation above, one finds that the calibration overlaps can be computed using the SE variables introduced above,
\begin{align}
    \label{eq:chap6-se-non-nishi-r}
    r &=  \int \dd{\eps} p_{\epsilon}(\eps) \dd{s_0}p_{s_0}(s_0) \int \dd{\w} \dd{\z} \cN(\z, \w ; \underline{0}, \Q) \,
    \sh\left(g_0\left( \z  , s_0,  \eps\right), \w , \V\right)^2 , \\
    \label{eq:chap6-se-non-nishi-rho}
    \nu &=  \int \dd{\eps} p_{\epsilon}(\eps) \dd{s_0}p_{s_0}(s_0) \int \dd{\w} \dd{\z} \cN(\z, \w ; \underline{0}, \Q) \,
    \sh\left(g_0\left( \z  , s_0,  \eps\right), \w , \V\right)  s_0 .
\end{align}

\paragraph{Performance analysis}
The mean squared error (MSE) on the reconstruction of $\X$ by the AMP algorithm is then predicted by 
\begin{gather}
\MSE(\X) = q - 2 m + q_0,
\end{gather}
where the scalar values used here correspond to the (unique) value of the diagonal elements of the corresponding overlap matrices. This MSE can be computed throughout the iterations of State Evolution. 
Similarly, the MSE in the reconstruction of the calibration variable can be computed as
\begin{gather}
    \MSE(\vect{s}) = r - 2 \nu + r_0,
\end{gather}
throughout the iterations.
Remarkably, the State Evolution MSEs follow precisely the MSE of the cal-AMP predictors along the iterations of the algorithm provided the procedures are initialized consistently. A random initialization of $\xh_i$ and $\sh$ in cal-AMP corresponds to an initialization of zero overlap $m = 0$, $\nu = 0$, with variance of the priors $q = q_0$, $r = r_0$, in the State Evolution.

\subsubsection{Bayes optimal State Evolution}
The SE equations can be greatly simplified in the Bayes optimal setting where the statistical model used by the student (priors $p_x$ and $p_s$, and channel $\pout$) matches the teacher model. 
In this case, the true unknown signal $\X_0$ is in some sense statistically equivalent to the estimate $\mat{\hat{X}}$ coming from the posterior. More precisely one can prove the Nishimori identities \cite{Opper1991, Iba1999, Nishimori2001} (or \cite{Kabashima2016} for a concise demonstration and discussion) implying that
\begin{gather}
    \q =  \mm, \quad  \V = \q_0  - \mm, \quad \qh  = \mh =\chih  \quad  \text{ and } \quad r = \nu.
\end{gather}
As a result the State Evolution reduces to a set of three equations
\begin{align}
    \label{eq:chap6-se-vect-glm-bayes-opt-r}
    r &=  \int \dd{\eps} p_{\epsilon}(\eps) \dd{s_0}p_{s_0}(s_0) \int \dd{\w} \dd{\z} \cN(\z, \w ; \underline{0}, \Q) \,
    \sh\left(g_0\left( \z  , s_0,  \eps\right), \w , \q_0 - \m\right)^2 , \\
    \label{eq:chap6-se-vect-glm-bayes-opt-m}
    \mm &= \int \dd{\x_0}p_{x_0}(\x_0) \int \D{\vect{\xi}} \vect{f}^x_1 \left( (\alpha \mh)^{-1}\left({\sqrt{\alpha \mh} \underline{\xi} + \alpha \mh\x_0}\right); (\alpha \mh)^{-1}  \right){\x_0}\T \, ,\\
    \label{eq:chap6-se-vect-glm-bayes-opt-mh}
    \mh  &= \int \dd{\eps} p_{\epsilon}(\eps) \dd{s_0}p_{s_0}(s_0) \int \dd{\w} \dd{\z} \cN(\z, \w ; \underline{0}, \Q)
    \gout\left(g_0\left( \z  , s_0,  \eps\right), \w , \q_0 - \m)\right) \times \\
    &\qquad \qquad \qquad \qquad \qquad \qquad \qquad \qquad  \qquad \qquad \qquad \gout\left(g_0\left( \z  , s_0,  \eps\right), \w , \q_0 - \m)\right)^T \notag,
\end{align}
with the block covariance matrix
\begin{gather}
    \label{eq:chap6-Q-bayes-opt}
    \Q = 
    \begin{bmatrix}
    \q_0 & \mm \\
    \\
    {\mm}\T & \mm \\ 
    \end{bmatrix}.
\end{gather}

\subsection{Replica free energy}
\label{sec:offline-fnrg}
The approximate inference via message passing presented in the previous sections is related to the replica approximation of the free energy of the problem defined as
\begin{gather}
    F_N = - \log p(\Y|\W) =  - \log \int \dd{\X} \dd{s} \pouts(\Y | \W\X, \vect{s}) p_x(\X) p_s(\vect{s}). 
\end{gather}
More precisely, the replica approximation \cite{Mezard1986} of the asymptotic expected free energy density
\begin{gather}
    f = - \lim_{N \to \infty} \frac 1 N \mathbb{E}_{\W}\int  \dd{\Y} p(\Y|\W) \log p(\Y|\W) 
\end{gather}
can be directly derived from GAMP on vector variables as presented in \cite{Aubin2018} (there applied to committee machines). With our notations, the Bayes optimal free energy density is
\begin{align}
    \label{eq:conjecture}
    - f =  \mathrm{max} \left\{ \; \mathrm{extr}_{\mm \, \mh} \left[ - \frac 1 2
  \mathrm{Tr}(\mm \, \mh)  + \mathcal{I}_x(\mh) + \alpha
  \mathcal{I}_z(\mm)\right]\right\}
\end{align}
with (again $\cal D$ is used for a standard multivariate Gaussian density) 
\begin{align}
    \mathcal{I}_x(\mh) &= \int_{\R^P} \D{\vect{\xi}} \int_{\R^P} \dd{\x_0} p_x(\x_0) e^{-\frac 1 2 \x_0\T \mh \x_0 + \vect{\xi}\T \mh^{1/2}\x_0} \log \left( \int_{\R^P} \dd{\x} p_x(\x) e^{-\frac 1 2 \x\T \mh \x + \vect{\xi}\T \mh^{1/2}\x}\right) , \\
    \mathcal{I}_z(\mm) &= \int_{\R^P} \dd{\y} \int_{\R^P} \D{\vect{\xi}} \int_{\R^P} \D{\z_0} \int \dd{s_0} p_s(s_0) \pout^s(\y | (\q_0 - \mm)^{1/2}\z_0 + \q^{1/2}\vect{\xi}; s_0)   \notag \\
    &\quad \quad  \quad \quad \quad \quad \quad \quad \quad \quad \quad \quad \quad\times\log \left( \int_{\R^P }\D{\z} \dd{s} p_s(s) \pout^s(\y | (\q_0 - \mm)^{1/2}\z + \q^{1/2}\vect{\xi}, s) \right).
\end{align}

In the case of mismatched teacher and student, the formula can be generalized and the expected free energy density
\begin{align}
    f &=  - \lim_{N \to \infty} \frac 1 N \mathbb{E}_{\W}\int \dd{\Y}\dd{\X_0} \dd{\vect{s}_0} \poutos(\Y | \W\X_0, \vect{s}_0) p_{x_0}(\X_0) p_{s_0}(\vect{s_0}) \nonumber\\
    &\qquad\qquad\qquad\qquad\qquad\qquad\qquad\qquad\times\log \left( \int \dd{\X} \dd{\vect{s}} \pouts(\Y | \W\X, \vect{s}) p_x(\X) p_s(\vect{s}) \right)
\end{align}
is now approximated as the extremum of a potential over all the overlap and auxiliary matrices 
\begin{gather}
    \q = \frac 1 N \sum_{i=1}^N \x_{a,i} \x_{a,i}\T \, , \quad 
    \mm = \frac 1 N \sum_{i=1}^N \x_{0,i} \x_{a,i}\T \, , \quad
    \q_{12} = \frac 1 N \sum_{i=1}^N \x_{a, i} \x_{b,i} \,,
\end{gather}
where $a$ and $b$ are indices of different replicas in the computation.
We obtain 
\begin{gather}
    - f 
    = \mathrm{max}\left\{  \; \mathrm{extr}_{\q, \qh, \mm, \mh, \q_{12}, \qh_{12}} \phi(\q,
    \qh, \mm, \mh, \q_{12}, \hat{\q}_{12})\, \right\},
\end{gather}
with
\begin{gather}
    \phi(\q, \qh, \mm, \mh, \q_{12}, \hat{\q}_{12}) = - \mathrm{Tr}( \mm \mh )+ \mathrm{Tr}( \q \qh) + \frac 1 2 \mathrm{Tr} (\q_{12} \qh_{12}) + \mathcal{I}_x(\qh,\mh,\qh_{12}) + \alpha \mathcal{I}_z(\q,\mm,\q_{12}) \, ,
\end{gather}
and
\begin{align}
    \mathcal{I}_x(\qh,\mh,\qh_{12}) &= \int \D{\vect{\xi}}  \int \dd{\x_0} p_x(\x_0) \log \left( \int \dd{\x} p_x(\x) e^{ \x_0\T\mh \x - \frac 1 2 \x\T(\qh + \qh_{12})\x + \vect{\xi}\T\qh_{12}^{1/2}\x}\right) \\
    \mathcal{I}_z(\q,\mm,\q_{12}) &= \int \dd{\y} \int \D{\vect{\xi}} \int \D{\z_0} \dd{\vect{s}_0} p_{\vect{s}_0}(\vect{s}_0) \poutos(\y | (\q_0 - \mm \q_{12}^{-1} \mm\T)^{1/2}\z_0 + \mm \q_{12}^{-1/2}\vect{\xi} ; \vect{s}_0) \notag \\
    &\quad \quad \quad \quad \quad \quad \quad  \quad \quad \quad \times  \log \left( \int \D{\z} \int \dd{\vect{s}} p_s(\vect{s}) \pouts(\y | (\q - \q_{12})^{1/2}\z + \q_{12}^{1/2}\vect{\xi}, \vect{s}) \right). 
\end{align}

Writing the extremization conditions for the potential $ \phi(\q, \qh, \mm, \mh, \q_{12}, \hat{\q}_{12})$ yields a set of self consistency equations for the overlap matrices. This set of coupled equations can be shown to be equivalent to the SE fixed point equations, \eqref{eq:chap6-se-vect-glm-bayes-opt-m}-\eqref{eq:chap6-se-vect-glm-bayes-opt-mh} for the Bayes optimal case, and \eqref{eq:chap6-se-non-nishi-qh}-\eqref{eq:chap6-se-non-nishi-q} for the mismatched case. This equivalence reveals the connection between the replica method and message passing algorithms at first apparently unrelated.

Finally, let us comment on the validity of the formula for the
asymptotic free energy in the Bayes optimal case. We conjecture that when one selects the maximum
value between all the extrema in \eqref{eq:conjecture}, the replica prediction is asymptotically
correct.  A rigorous justification can be provided using the approach
used for the committee machine in \cite{Aubin2018}, based on
interpolation technics developed in
\cite{Barbier2017a,Barbier2019}
. It is rather
straightforward to repeat these steps for the present problem, which
should lead to a complete proof of the conjecture \eqref{eq:conjecture}. 

\begin{algorithm*}[t]
\caption{Offline Gain Calibration State Evolution\label{alg:chap6-offline-cal-se}}   
\begin{algorithmic}
    \State {\bfseries Input:} matrix $\Y \in \R^{M \times P}$ and matrix $\W \in \R^{M \times N}$:
    \vspace{0.2cm}
    \State \emph{Initialize:} 
    \vspace{0.1cm}
    \State  $t=0$, $m^{(0)}= 0$,  $V^{(0)}= q_0 = \rho $, 
    \State $\forall\ \mu =1, \cdots, N_{\rm MC} \ \  s_{0,\mu} \sim p_{s_0}(s_{0,\mu}) $
    \vspace{0.1cm}
    \Repeat   
    \Statex \hspace{0.35cm} 1) Draw Monte Carlo samples for update of $\hat{m}$ \eqref{eq:chap6-cal-se-bo-mh} and $r$ \eqref{eq:chap6-cal-se-bo-r} 
    \Statex $\qquad \forall \ \mu = 1, \cdots, N_{\rm MC}$
        \vspace{-0.3cm}
    \begin{gather*}
        \begin{array}{ll}
            \begin{array}{l}
            \quad \z_{\mu}, \w_{\mu} \sim \cN(\z_{\mu}, \w_{\mu} ; \underline{0}, \Q^{(t)}) \\
            \quad \eps \sim p_\epsilon(\eps) \\
            \quad \y_{\mu} = g_0(\z_{\mu}, s_{0,\mu}, \eps)
            \end{array}
        & 
        \quad \text{ with } \quad
        \Q_{(k)}^{(t)} = 
        \begin{bmatrix}
        q_0I_P & m^{(t)}I_P\\
        \\
        m^{(t)}I_P & m^{(t)}I_P\\
        \end{bmatrix}         \in R^{2P\times 2P}
\end{array}
    \end{gather*} 
    \vspace{-0.3cm}
    \Statex \hspace{0.35cm} 2) Compute integrands
    \vspace{0.1cm}
    \Statex \hspace{0.85cm} 2.2) Compute $\sh_\mu(\y_\mu, \z_\mu, \w_\mu, \V^{(t)})$  \eqref{eq:chap6-gain-sh}
    \Statex \hspace{0.85cm} 2.3) Compute  $\gout_{\mu}(\y_\mu, \z_\mu, \w_\mu, \V^{(t)})$ \eqref{eq:chap6-gain-gout}
    \vspace{0.1cm}
    \Statex \hspace{0.35cm} 3) Update $r$ and $\hat{m}$
    \begin{align*}
        r^{(t)}= \frac{1}{N_{\rm MC}}\sum_{\mu =1}^{N_{\rm MC}} \left(\sh^{(t)}_\mu\right)^2
    \qquad
        \hat{m}^{(t)}= \frac{1}{N_{\rm MC}}\sum_{\mu =1}^{N_{\rm MC}} \left(\gouts_{\mu}\right)^2
    \end{align*}

    \Statex \hspace{0.35cm} 3) Update $m^{(t+1)}$ by numerical integration using \eqref{eq:chap6-cal-se-bo-m}, and $V^{(t+1)}= q_0 - m^{(t+1)}$. 
    \vspace{0.1cm}
    \State $t = t + 1$
    \vspace{0.1cm}
    \vspace{0.05cm}
    \Until{convergence} 
    \State {\bfseries Output:} time series $\{V^{(t)}, m^{(t)}, \hat{m}^{(t)} \; ; \; t = 1 \cdots t_{\rm max} \}$
\end{algorithmic}
\end{algorithm*}

\section{Online algorithm and analysis}
\label{sec:chap6-online}
In machine learning applications, the \emph{offline} strategy, where all the training data is exploited at once, is sometimes made difficult by its memory cost. Instead, data may be partitioned into mini-batches of a few samples, or even considered one sample at the time. This approach may be preferred even when memory is not an issue, as in deep learning with Stochastic Gradient Descent (SGD) \cite{Bottou2018}. 
Here we will be interested in the Bayesian online learning of the calibration variables as the observations are received. The main concept behind Bayesian online learning algorithms, such as Assumed Density Filtering (ADF) \cite{Opper1999, Minka2001}, is to update a prior along upon receiving/processing different data points.  
In the statistical physics literature, Bayesian online learning algorithms were proposed and analyzed for simple neural networks 
\cite{Kinouchi1992, Biehl1994, Solla1999, Saad1999a} and compressed sensing \cite{Rossi2016}.
 
In the present section, we propose an online algorithm, online cal-AMP, for the calibration problem of the GLM. Our algorithm takes inspiration from the streaming version of AMP for the simple GLM of \cite{Manoel2018}, which we review in the following section. We also derive the corresponding State Evolution analysis.

\subsection{Streaming AMP for online reconstruction in the GLM}
\label{sec:chap3-streaming-amp}

In \cite{Manoel2018}, a mini-batch version of the GAMP algorithm is proposed. On the example of the GLM, one imagines receiving at each iteration $k$ a subset  $\y\kk$ of the components of $\y \in R^M$ generated via a channel
\begin{gather}
    \y \sim \pout(\y | \W\x) \, ,
\end{gather}
from which one wishes to reconstruct $\x \in \R^N$. Bayes formula gives the posterior distribution over $\x$ after seeing $k$ mini-batches
\begin{align}
    p(\x|\y\kk, \{\y_{(k-1)}, \cdots \y_{(1)}\},\W) = \frac{p(\y\kk|\x,\W)p(\x|\{\y_{(k-1)}, \cdots \y_{(1)}\},\W)}{\int \dd{\x}p(\y\kk|\x,\W)p(\x|\{\y_{(k-1)}, \cdots \y_{(1)}\},\W)} .
\end{align}
The formula suggests the iterative scheme of using as a prior on $\x$ at iteration $k$ the posterior obtained at iteration $k-1$. This idea can be implemented in different approximate inference algorithms, as proposed by \cite{Broderick2013} using a variational method. In the regular version of GAMP \cite{Rangan2011} an effective factorized posterior is given at convergence by the input update functions (scalar equivalent of \eqref{eq:chap6-vect-amp-f1x}-\eqref{eq:chap6-vect-amp-f2x}):
\begin{align}
p(\x|\y, \W) \simeq \prod_{i=1}^N \frac{1}{\cZ_x(\lambda_i, \sigma_i)}p_x(x_i)\exp\Big\{-\frac{(\lambda_i-x_i)^2}{2 \sigma_i}\Big\}.     
\end{align}
Plugging this posterior approximation in the iterative scheme yields the mini-AMP algorithm using the converged values of ${\lambda_{(\ell),i}}$ and ${\sigma_{(\ell),i}}$ at each anterior mini-batch $\ell < k$ to compute the prior
\begin{align}
    p_{x}^{(k)}(\x) = p(\x|\{\y_{(k-1)}, \cdots \y_{(1)}\}, \W) \simeq \prod_{i=1}^N \frac{1}{\cZ_{x,i}} \; p_x(x_i) \exp\Big\{-\sum\limits_{\ell=1}^{k-1}\frac{(\lambda_{(\ell), i}-x_i)^2}{2 \sigma_{(\ell), i}}\Big\},     
\end{align}
where the $\cZ_{x,i}$ normalizes its respective marginal.
Compared to a naive mean-field variational approximation of the posterior, AMP takes into account more correlations and is indeed found to perform better in experiments reported by \cite{Manoel2018}. Another advantage of the AMP based online inference is that it is amenable to theoretical analysis by a corresponding State Evolution. 
In the following we apply these ideas to the online learning of the calibration variables in the model we defined in \citesec~\ref{sec:chap6-model}.

\subsection{Online cal-AMP}
We consider here the analysis of the online reconstruction of the calibration variable $\vect{s}_0$ by tracking $\vect{s}$ as the observations $\y\kk \in \R^M$ are treated consecutively. 
We readily adapt the strategy of \cite{Manoel2018},
by using at step $k + 1$ the approximate posterior on $\vect{s}$ at step $k$ as an effective prior 
on $s_\mu$ in cal-AMP.

\paragraph{Algorithm}
The AMP algorithm consists here in restarting cal-AMP at each new observation $\y\kk$ (that is with $P=1$ sample in the notation of \citesec~\ref{sec:chap6-offline}), while updating the prior used for the calibration variable.
From the definition of the approximate posterior \eqref{eq:chap6-vect-amp-post-s}, we obtain the recursion on the effective prior ${p_{s_\mu}^{(k+1)}}$ on $s_\mu$:
\begin{align}
    \label{eq:chap6-online-rec}
    {p_{s_\mu}^{(k+1)}}(s_\mu) & = m^{s, (k)}_\mu(s_\mu) \\
    & = \frac 1 {\Zout^{(k)}} \int \dd{z} \pouts \left(y_{(k), \mu} | z, s_\mu \right){p_{s_\mu}^{(k)}}(s_\mu) \cN(z; \omega_{(k), \mu}, V_{(k), \mu} ),
\end{align}  
where the output variables $\omega_{(k), \mu}$ and $V_{(k), \mu}$ correspond to the values at convergence (or at the last iteration $t_{\rm max}$) of the cal-AMP algorithm at the previous step $k$. 
In \citesec~\ref{sec:chap6-gain-cal}, we consider the gain calibration problem and specify effective strategies to implement this recursion within the AMP algorithm. 

\paragraph{State Evolution}
The streaming State Evolution analysis of the online cal-AMP is also adapted using the above recursion. Note that the effective prior at a given step/sample $P$ (we assume the samples to be processed one at a time) depends on the output variables of the algorithm for all the previously seen samples. Expending the recursion above we have:
\begin{gather}
    \label{eq:chap6-online-rec-dvp}
    {p_{s_\mu}^{(P)}}(s_\mu) = \frac 1 {\Zout^{(P-1)}} \int \prod_{k=1}^{P-1} \Big( \dd{z_{(k)}}   \pouts \left(y_{(k), \mu} | z_{(k)}, s_\mu \right) \cN(z_{(k)}; \omega_{(k),\mu}, V_{(k), \mu} ) \Big) p_{s}(s_\mu)
\end{gather}
with
\begin{gather}
    \label{eq:chap6-online-zout}
    \Zout^{(P)} = \Zout^{(P)}(\{y_{(k), \mu}, \omega_{(k), \mu}, V_{(k), \mu} \}_{k=1}^P) \, ,
\end{gather}
where again for each $k$ the output variables $\vect{\omega}_{(k)}$ and $\vect{V}_{(k)}$ are the converged values for the corresponding step $k$. The dependence of the normalization $\Zout^{(P)}$ on the output variables 
is reflected in the definitions at step $P$ of the output update function
$\vect{\gouts}^{(P)} = \partial \log \Zout^{(P)} / \partial \vect{\omega}_{(P)}$ and calibration update function $\vect{\sh}^{(P)} = \int \dd{\vect{s}} \vect{s} \, m^{s, (P)}(\vect{s})$. Therefore, the State Evolution involves an averaging over all the output variables $\vect{\omega}_{(k)}$ and $\vect{V}_{(k)}$ relative to the already processed samples.  

More precisely, in the Bayes optimal setting, it is easy to see from the cal-AMP SE, that the online algorithm is characterized by the equations:
\begin{align}
    \label{eq:chap6-cal-se-bo-r}
    r_{(P)} &=  \int \dd{\epsilon} p_{\epsilon}(\epsilon) \dd{s_0}p_{s_0}(s_0) \int \dd{\omega} \dd{z} \cN(z, \omega ; \underline{0}, \Q_{(P)}) \, \\
    &\qquad \qquad  \prod_{k=1}^{P-1} \int \dd{\omega\kk} \dd{z\kk} \cN(z\kk, \omega\kk ; \underline{0}, \Q\kk) \;
    \sh^{(P)}\left(g_0\left( z  , s_0,  \epsilon  \right), \omega , q_0 - m\right)^2 , \notag \\
    \label{eq:chap6-cal-se-bo-m}
    m_{(P)}&= \int \dd{x_0}p_{x_0}(x_0) \int \D{\xi} f^x_1 \left( (\alpha \hat{\chi})^{-1}
     \left({\sqrt{\alpha \hat{m}} \xi + \alpha \hat{m}_{(P)} x_0}\right); (\alpha \hat{m}_{(P)})^{-1}  
    \right){x_0} \, , \\
    \hat{m}_{(P)}  &= \int \dd{\epsilon} p_{\epsilon}(\epsilon) \dd{s_0}p_{s_0}(s_0) \int \dd{\omega} \dd{z} \cN(z, \omega ; \underline{0}, \Q_{(P)}) \; \notag\\
    &\qquad \qquad  \prod_{k=1}^{P-1} \int \dd{\omega\kk} \dd{z\kk} \cN(z\kk, \omega\kk ; \underline{0}, \Q\kk) \;
    \gouts^{(P)}\left(g_0\left( z  , s_0,  \epsilon\right), \omega , q_0 - m\right)^2,
    \label{eq:chap6-cal-se-bo-mh}
\end{align}
where the $2\times2$ covariance matrices $\Q\kk$, are analogous to the block covariance matrix \eqref{eq:chap6-Q-bayes-opt} for each step $k$ using the corresponding scalar fixed points $m\kk$.
In the following \citesec, we will discuss how to implement this State Evolution in practice, focusing on the specific problem of gain calibration.

\begin{algorithm*}[h!]
\caption{Online Gain Calibration State Evolution\label{alg:chap6-online-cal-se}}
\begin{algorithmic}
    \State {\bfseries Input:} matrix $\Y \in \R^{M \times P}$ and matrix $\W \in \R^{M \times N}$:
    \State \emph{Initialize:} $t = 0$, 
        \State $\forall \ \mu = 1, \cdots, N_{\rm MC}$
        \vspace{-0.3cm}
        \begin{gather*}
            s^0_\mu \sim p_{s^0}(s^0_\mu) \\
            \Lambda_{0,\mu}=0 \, , \quad \Sigma_{0, \mu}=0
        \end{gather*} 
    \For{$k = 1, \cdots, P$}
        \State \emph{Initialize}: $t=0$, $m^{(0)}_{(k)} = 0$,  $V^{(0)}_{(k)} = q_0 = \rho $
        \Repeat   
        \Statex \hspace{0.35cm} 1) Draw Monte Carlo samples for update of $\hat{m}$ \eqref{eq:chap6-cal-se-bo-mh} and $r$ \eqref{eq:chap6-cal-se-bo-r} 

        \Statex $\qquad \forall \ \mu = 1, \cdots, N_{\rm MC}$
         \vspace{-0.3cm}
        \begin{gather*}
            \begin{array}{ll}
                \begin{array}{l}
                \quad (z_{\mu, k}, \omega_{\mu, k}) \sim \cN(z_{\mu, k}, \omega_{\mu, k} ; \underline{0}, \Q^{(t)}_{(k)}) \\
                \quad \epsilon \sim p_\epsilon(\epsilon) \\
                \quad y_{\mu, k} = g^0(z_{\mu, k}, s^0_\mu, \epsilon)
                \end{array}
            & 
            \quad \text{ with } \quad
            \Q_{(k)}^{(t)} = 
            \begin{bmatrix}
            q_0 & m^{(t)}_{(k)} \\
            \\
            m^{(t)}_{(k)}& m^{(t)}_{(k)} \\
            \end{bmatrix}         
    \end{array}
        \end{gather*} 
        \vspace{-0.3cm}
        \Statex \hspace{0.35cm} 2) Following step (2) of \citealg~\ref{alg:chap6-online-cal-amp} with
        \Statex \hspace{0.75cm} - samples of previous steps $\{y_{\mu, l}, z_{\mu, l}, \omega_{\mu, l}\}_{l\leq k-1}$ at convergence 
        \Statex \hspace{0.75cm} -  current $y_{\mu, k}, z_{\mu, k}, \omega_{\mu, k}$
        \Statex \hspace{0.75cm} -  current $V_{(k)}^{(t)}$
        \vspace{0.1cm}
        \Statex \hspace{0.85cm} 2.1) Update $\Lambda_{k, \mu},\Sigma_{k, \mu}$ 
        \Statex \hspace{0.85cm} 2.2) Compute $\sh_\mu$ 
        \Statex \hspace{0.85cm} 2.3) Compute  $\gouts_{\mu,k}$ 
        \vspace{0.1cm}
        \Statex \hspace{0.35cm} 3) Update $r_{(k)}$ and $\hat{m}_{(k)}$
        \begin{align*}
            r^{(t)}_{(k)} = \frac{1}{N_{\rm MC}}\sum_{\mu =1}^{N_{\rm MC}} \left(\sh^{(t)}_\mu\right)^2
        \qquad
            \hat{m}^{(t)}_{(k)} = \frac{1}{N_{\rm MC}}\sum_{\mu =1}^{N_{\rm MC}} \left(\gouts_{\mu,k}\right)^2
        \end{align*}
    
        \Statex \hspace{0.35cm} 4) Update $m^{(t+1)}_{(k)}$ by numerical integration using \eqref{eq:chap6-cal-se-bo-m}, and $V^{(t+1)}_{(k)} = q_0 - m^{(t+1)}_{(k)}$. 
        \vspace{0.1cm}
        \Statex \hspace{0.35cm} $t = t+1$
        \vspace{0.15cm}
        \Until{convergence} 
    \EndFor
    \State {\bfseries Output:} time series $\{V\kk^{(t)}, m\kk^{(t)}, \hat{m}\kk^{(t)} \; ; \; t = 1, \cdots, t_{\rm max} \}_{k=1}^P $
\end{algorithmic}
\end{algorithm*}

\begin{figure}[t]
    \centering
    {\includegraphics[width=0.9\textwidth]{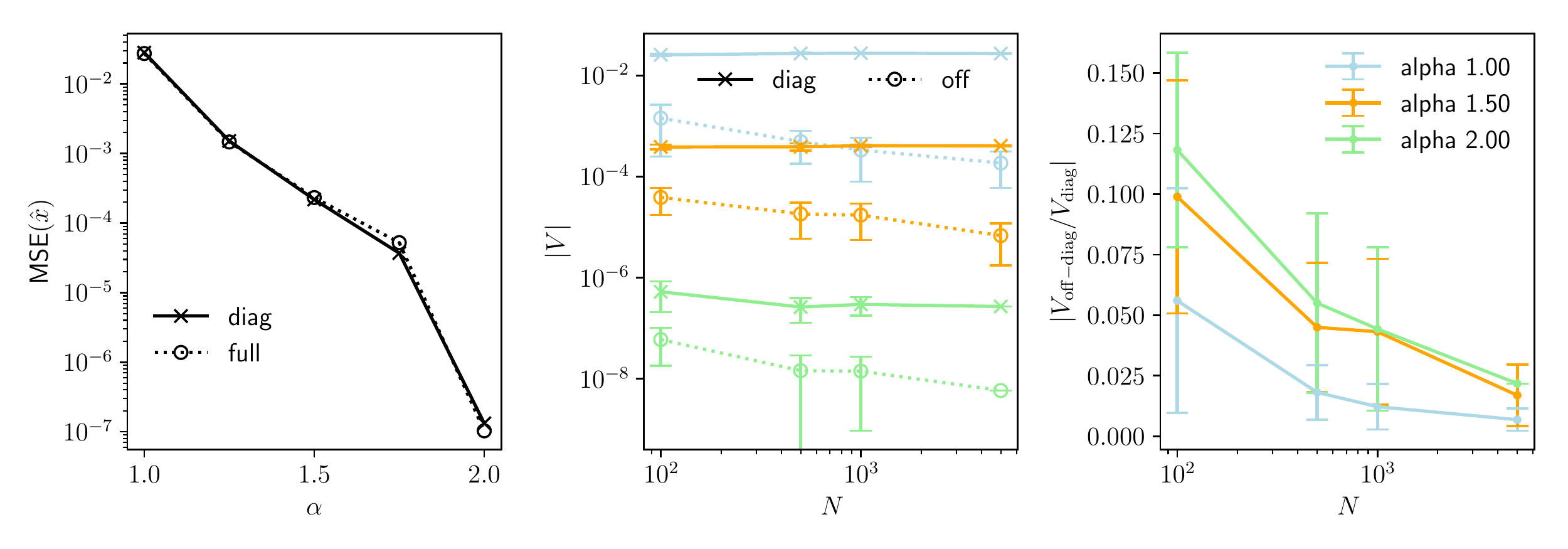}
    \label{fig:chap6-calamp-mul-ansatz-P2}}
    {\includegraphics[width=0.9\textwidth]{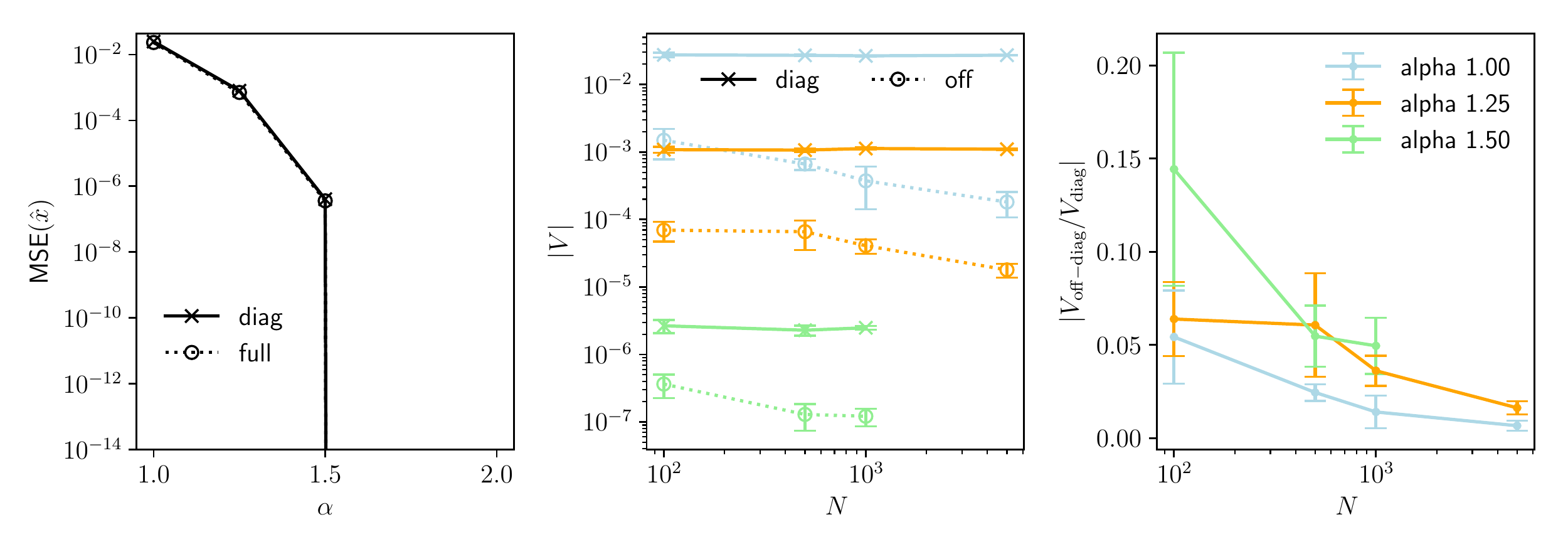}
    \label{fig:chap6-calamp-mul-ansatz-P3}}
    \caption{Full covariance matrices and departure from the diagonal ansatz for Gaussian inputs $\rho=1$ \textbf{(top row)} $P=2$ \textbf{(bottom row)} $P=3$. \textbf{Left:} Comparison of the MSEs achieved by cal-AMP for full unconstrained covariance matrices and diagonal covariance matrices for $N=1000$. \textbf{Middle:} Average value at fixed points of the diagonal elements and the off-diagonal elements of $\V$ when unconstrained as a function of the size of the problem $N$. \textbf{Right:} Average absolute value of the ratio between off-diagonal and diagonal elements of $\V$ as a function of $N$.
      \label{fig:chap6-calamp-mul-ansatz}}
\end{figure}

\section{Numerical tests}
\label{sec:chap6-numerics}

\subsection{Gain calibration setting and update functions}
\label{sec:chap6-gain-cal}
As a test case, we consider the following problem of gain calibration \cite{Gribonval2012,Shen2013,Schulke2013,Schulke2015}. The input signal is known to be $\rho$-sparse and distributed according to a Gauss-Bernoulli distribution, 
\begin{align}
    \forall \,  i = 1, \cdots, N \, , \quad & p_{x_0}(\x_i) = \prod_{k=1}^P \left( \rho \, \cN(x_{(k),i}; 0, 1) + (1 - \rho) \,\dirac(x_{(k),i}) \right) \, . 
\end{align}
Each component of the output includes a division by a calibration variable that is uniformly distributed in a positive interval $[a,b]$:
\begin{align}
    \forall\ \mu = 1, \cdots, M \, , \quad p_{s_0}(s_\mu) &= \mathds{1}_{[a,b]} / (b -a)  \, , \quad 0<a<b, \\
     \y_\mu &= \frac{1}{s_{0,\mu}}(\vect{w}_\mu\T \X_0 + \eps) \, , \quad \eps \sim  \cN(\eps, \vect{0}, \Delta I_P) \,. 
\end{align}
We will consider both the offline reconstruction where the $P$ samples are exploited simultaneously and the online case. 

\paragraph{Output functions} For any value of $P$, the output functions have analytical expressions in this setting. The channel distribution and partition function are
\begin{align}
    \pout(\y_\mu|\z_\mu) &= \int_a^b \dd{s_\mu} p_s(s_\mu)(s_\mu)^P\cN(\z_\mu; s_\mu \y_\mu, \Delta I_P), \\
    \Zout(\y,\w,\V) &= \int \dd{\z} \pout(\y|\z)\cN(\z; \w, \V) = \int \dd{\z}\int_a^b \dd{s} p_s(s) (s)^P\cN(\z; s \y, \Delta I_P) \cN(\z; \w, \V) \notag\\
                &= \int_a^b \dd{s} p_s(s) (s)^P\cN(\w; s \y, \V + \Delta I_P)
\end{align}
which give
\begin{align}
    \label{eq:chap6-gain-gout}
    \gout(\y,\w,\V) &= \frac 1 \Zout \partial_{\w}\Zout = (\V + \Delta I_P)^{-1}(\sh(\y,\w,\V)\y - \w), \\
    \partial_{\w} \gout(\y,\w,\V) &= \Cs(\y,\w,\V) (\V + \Delta I_P)^{-1} \y \y\T(\V + \Delta I_P)^{-1} - (\V + \Delta I_P)^{-1}.
\end{align}
These expressions involve the estimate and variance of the calibration variables under the approximate posterior which can be computed as
\begin{align}
    \label{eq:chap6-gain-sh}
    \sh(\y,\w,\V) &= f_1^s(\y,\w,\V) = \frac{\int_a^b \dd{s} (s)^{P+1} \cN(\z; s \y, \V + \Delta I_P) }{\int_a^b \dd{s} (s)^P\cN(\z; s \y, \V + \Delta I_P) } = \frac{\mathcal{I}(P+1, \ms, \sigs, a, b)}{\mathcal{I}(P, \ms, \sigs, a, b)}, \\
    \label{eq:chap6-gain-Cs}
    \Cs(\y,\w,\V) &= f_2^s(\y,\w,\V) = \frac{\mathcal{I}(P+2, \ms, \sigs, a, b)}{\mathcal{I}(P, \ms, \sigs, a, b)} - \sh(\y,\w,\V)^2,
\end{align} 
where
\begin{align}
        \mathcal{I}(P, \ms, \sigs, a, b) &= \int_a^b \dd{s} s^P e^{-\frac 1 {2\sigs} (s-\ms)^2 } \, , \\
        \sigs &= (\y\T (\V + \Delta I_P)^{-1} \y)^{-1} \, , \\
        \sigs^{-1}\ms &= \y\T(\V + \Delta I_P)^{-1} \w
\end{align}
and $\mathcal{I}(P, \ms, \sigs, a, b)$ can be computed using gamma functions as explained in \cite{Schulke2013,Schulke2015}.

\paragraph{Input functions}
For a Gauss-Bernoulli prior on the entries of $\X$, and assuming the AMP variances $\sig_i$ are diagonal matrices, the input update functions can be written component-wise with scalar arguments:
\begin{align}
    f_{1}^x(\lambda, \sigma) & 
     = 
        \left.\left(
         \rho \frac{ \lambda}{(1 + \sigma )^{3/2}} e^{-\frac{\lambda^2}{2(1 + \sigma)}}
        \right) \middle/\left(
         \rho \frac{e^{-\frac{\lambda^2}{2(1 + \sigma)}}}{(1 + \sigma )^{1/2}}  + ( 1 - \rho) \frac{e^{-\frac{\lambda^2}{2 \sigma}}}{\sigma^{1/2}}
        \right) \right.
        ,
\end{align}
and
\begin{align}
    f_{2}^x(\lambda, \sigma)
    = 
    \left.
    \left(    
    \rho \frac{\sigma(1 + \sigma) + (\lambda)^2}{(1 + \sigma )^{5/2}} e^{-\frac{\lambda^2}{2(1 + \sigma)}}
    \right) \middle/ \left(
    \rho \frac{e^{-\frac{\lambda^2}{2(1 + \sigma)}}}{(1 + \sigma )^{1/2}}  + ( 1 - \rho) \frac{e^{-\frac{\lambda^2}{2 \sigma}}}{\sigma^{1/2}}
    \right)\right.
    - {f_{1, k}^x}^2,
\end{align}
so that $\hat{x}_{i,k} = f_1^x(\lambda_{i,k}, \sigma_{i,kk})$ and $C^x_{i,kk} = f_2^x((\lambda_{i,k}, \sigma_{i,kk})$. 

\subsection{Offline tests}
 
The offline AMP algorithm is directly given by \citealg~\ref{alg:chap6-vect-amp} replacing the values of the output and input functions presented in the previous paragraph. In our numerical tests, we additionally
impose that the covariance matrices $\V_\mu$, $\dgout_\mu$ and $\sig_i$ are diagonal. This assumption lightens the numerics when considering a larger number of samples $P$.

Consequently, we first investigate how the algorithm behaves under a diagonal ansatz for the covariances. On \citefig~\ref{fig:chap6-calamp-mul-ansatz}, we show that the performance of the algorithm with diagonal covariances is comparable to the performance of the algorithm with a full covariance matrix in terms of signal reconstruction. Furthermore, we find that the off-diagonal elements tend to vanish as we increase the size of the problem, and conjecture that they are null in the thermodynamic limit. In the following, we adopt the diagonal ansatz for covariances.

Under the assumption of diagonal covariance matrices, the State Evolution equations \eqref{eq:chap6-se-vect-glm-bayes-opt-m}, \eqref{eq:chap6-se-vect-glm-bayes-opt-mh} involve $P \times P$ matrices proportional to the identity (given the $P$ examples are statistically equivalent). Therefore they can be parametrized with one scalar value, the unique diagonal element, noted respectively $m$ and $\hat{m}$. While the update of $m$ only requires a two dimensional integral that can be performed numerically, we resort to a Monte Carlo sampling for the update of $\hat{m}$. The procedure is described in \citealg~\ref{alg:chap6-offline-cal-se}.

On the top row of \citefig~\ref{fig:chap6-calamp-mul-diag-x} and respectively of \citefig~\ref{fig:chap6-calamp-mul-diag-s}, we report the reconstruction performance by the cal-AMP algorithm of the signal $\X_0$ and the calibration variable $\vect{s}_0$,
in the plane $\rho-\alpha$, for different values of $P$. These phase diagrams  are similar to the ones reported in \cite{Schulke2013, Schulke2015} and will be compared with the online case described in the next \citesec. 
The performances are compared with the information theoretic threshold $\alpha_{\rm min}$ corresponding to the minimal number of observations necessary to reconstruct the unknown signal and calibration variables,
\begin{gather}
    M_{\rm min} \times P = \rho N \times P + M_{\rm min} \Rightarrow \alpha_{\rm min} = \frac{M_{\rm min}}{ N} = \rho\frac{P}{P-1} \, ,
\end{gather}
and the sharp threshold of reconstruction of the fully calibrated Bayesian AMP $\alpha_{\rm CS}$ \cite{Krzakala2012}.

\begin{figure}[t]
    \centering
    \subfloat[Offline]
    {\includegraphics[width=0.35\textwidth]{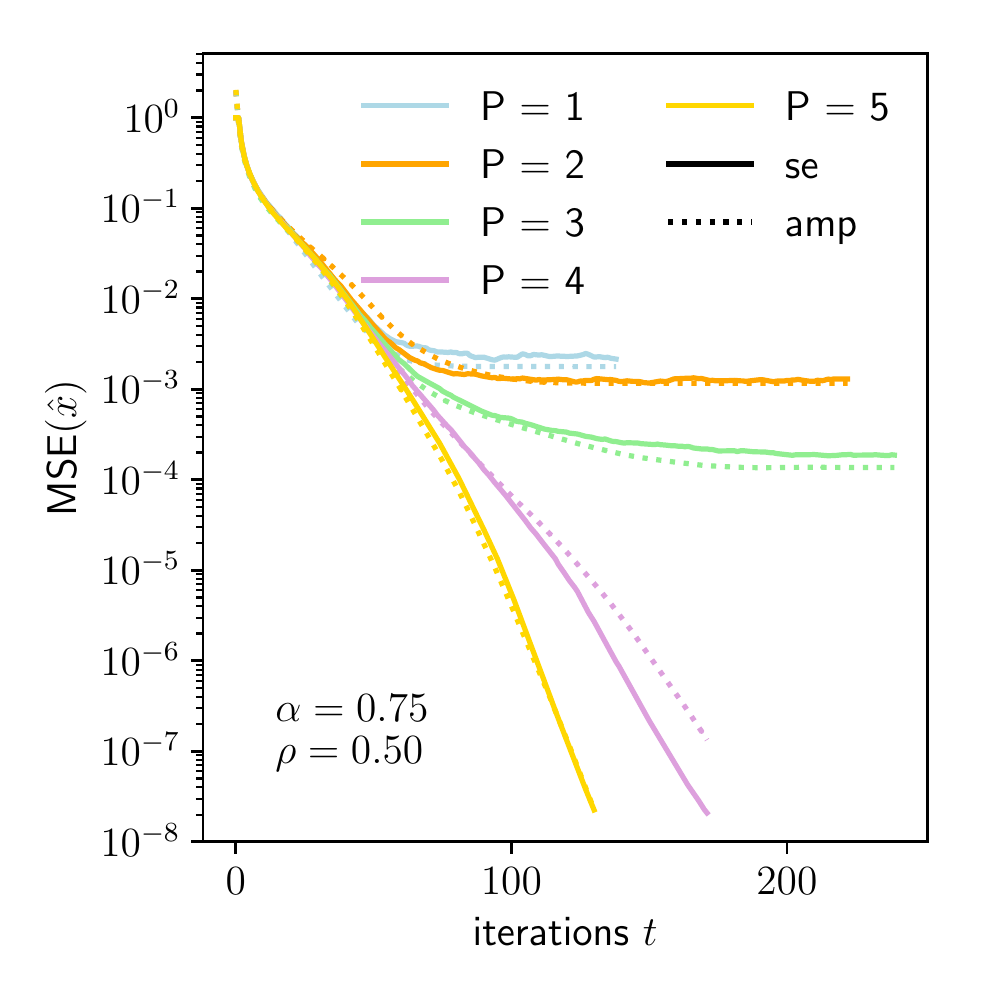}
    \label{fig:chap6-calamp-mul-dyn-offline}}
    \hspace{0.05\textwidth}
    \subfloat[Offline]
    {\includegraphics[width=0.35\textwidth]{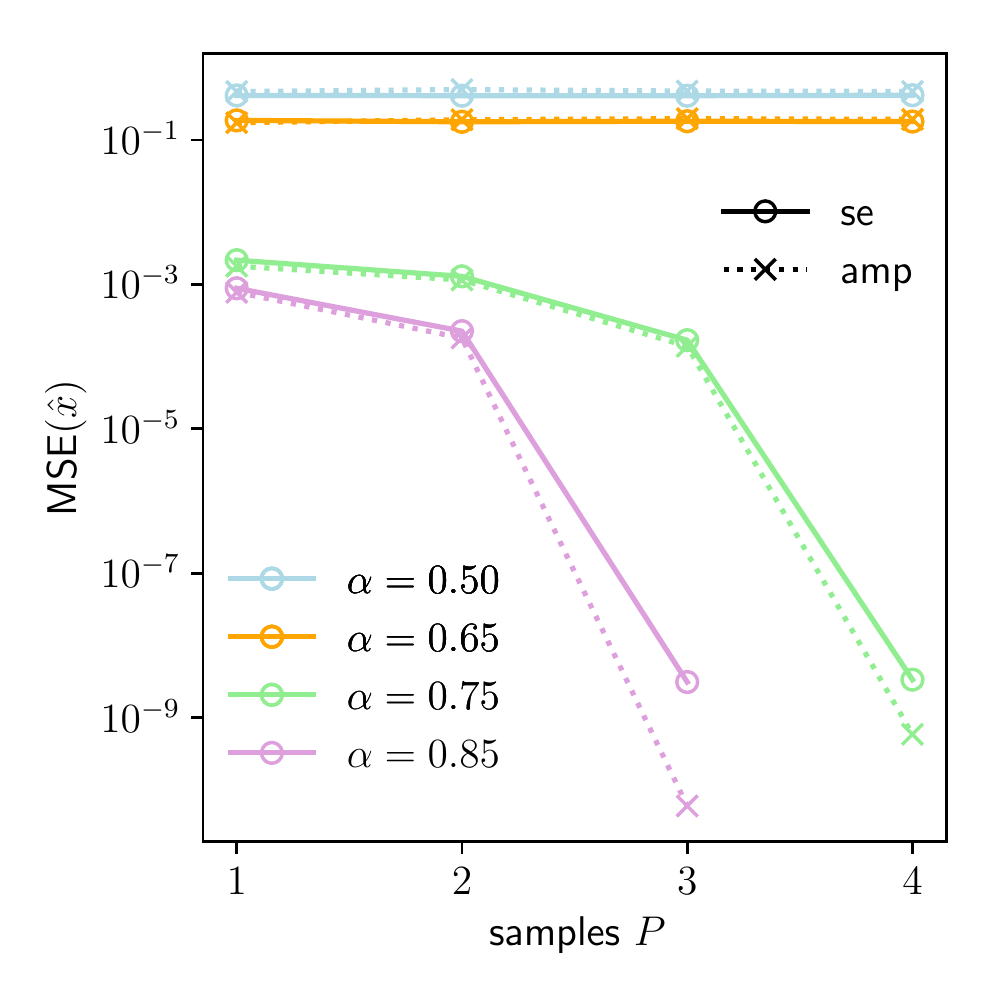}
    \label{fig:chap6-calamp-mul-comp-offline}} \\
    \subfloat[Online]
    {\includegraphics[width=0.35\textwidth]{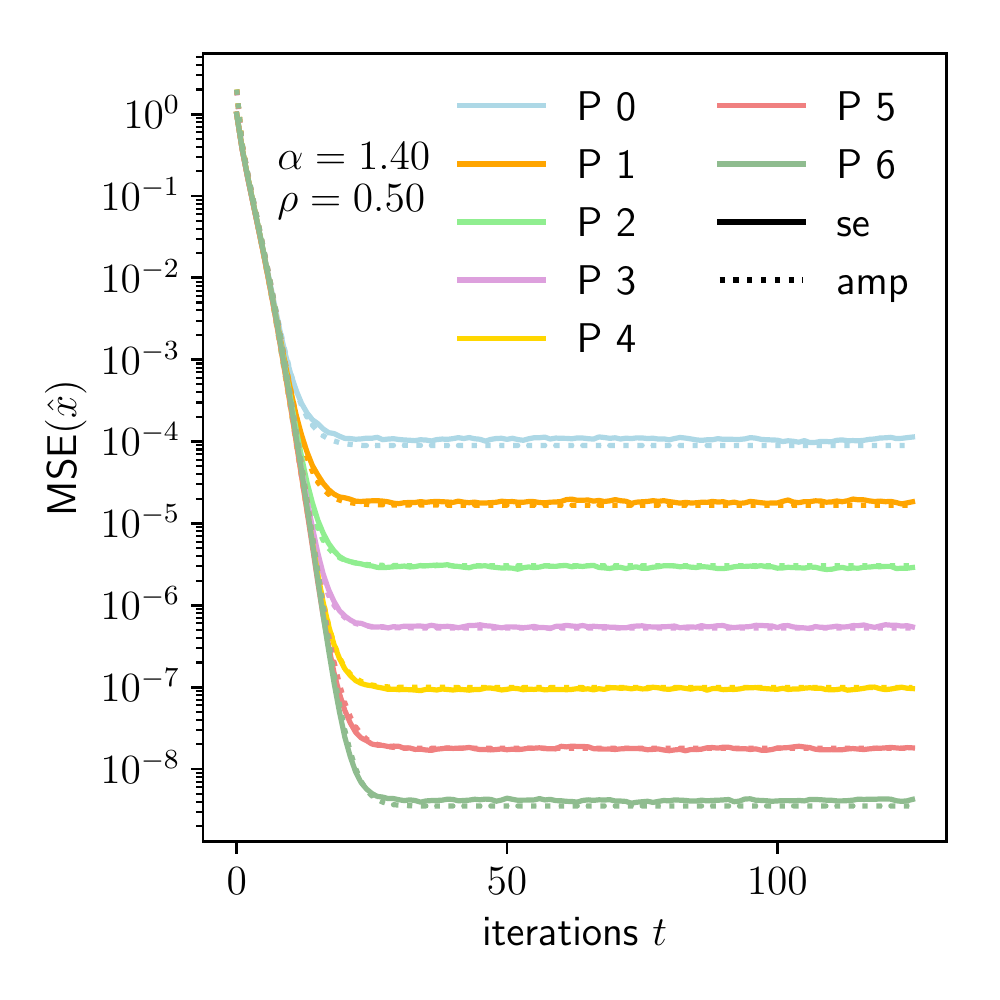}
    \label{fig:chap6-calamp-mul-dyn-online}}
    \hspace{0.05\textwidth}
    \subfloat[Online]
    {\includegraphics[width=0.35\textwidth]{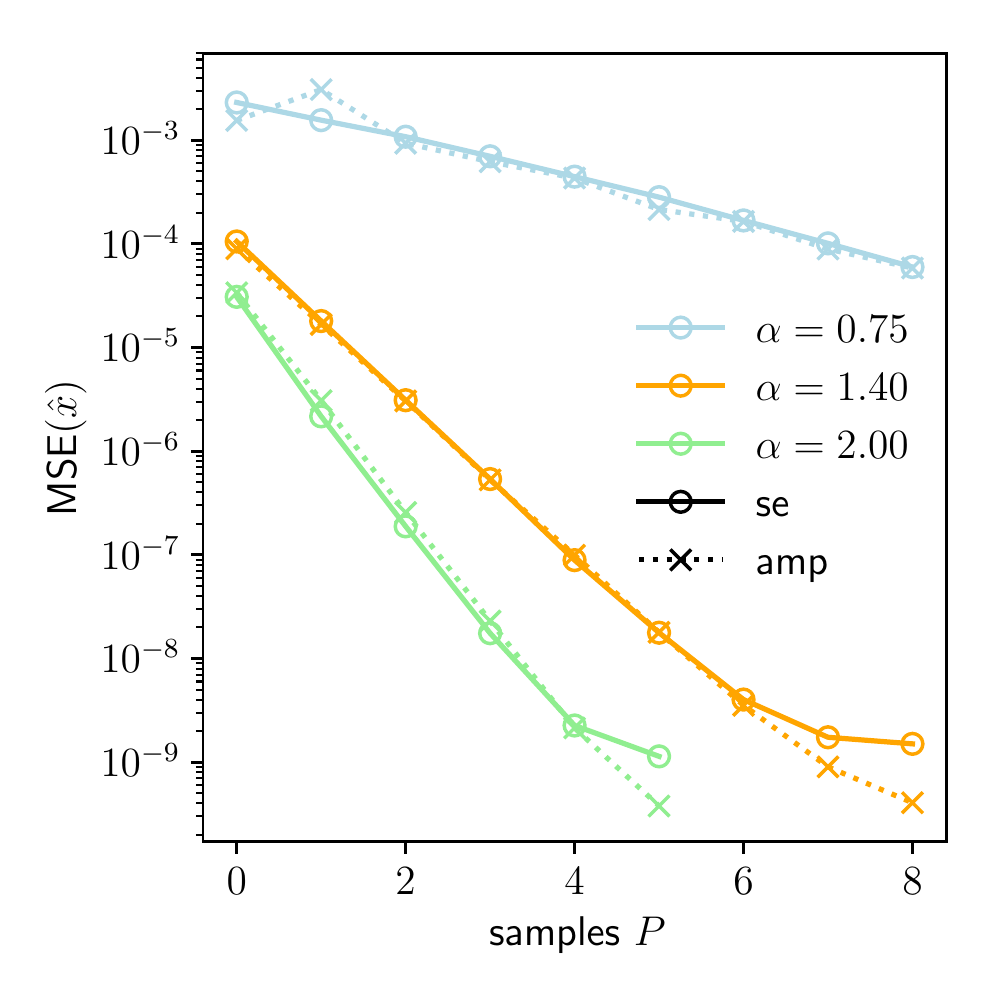}
    \label{fig:chap6-calamp-mul-comp-online}}
    \caption{Comparison of mean squared errors in reconstructing the signal $\X_0$ using cal-AMP and predicted by SE for the gain calibration problem with calibration variables uniformly distributed in $[0.95,1.05]$ and $N=10^4$. Along the iterations of the algorithm and comparing fixed points we find a very good agreement of the two procedures. Nevertheless, the Monte Carlo integration in the SE does not allow for great numerical precision for low errors. We start seeing discrepancies between SE and AMP for errors below $10^{-8}$.}
\end{figure}

On \citefig~\ref{fig:chap6-calamp-mul-dyn-offline}, we check numerically that the derived State Evolution predicts the behavior of the AMP algorithm for gain calibration. The MSEs of the two procedures are indeed consistent along the iterations of the algorithm. On \citefig~\ref{fig:chap6-calamp-mul-comp-offline}, we also report an almost perfect agreement of the fixed points of SE and cal-AMP in terms of the MSE of the cal-AMP estimate of $\X_0$ as we vary the number of samples $P$.

\subsection{Online tests}
The online algorithms include supplementary operations to update the effective prior on the calibration variable from one step to the next. In this setting the recursion is 
\begin{align}
    {p_{s_\mu}^{(k+1)}}(s_\mu) 
    &= \frac 1 {\Zout^{(k+1)}} 
    \int \dd{z}
    \cN(\y_{(k), \mu}; z/ s_\mu, \Delta ) \cN(z; w_{(k), \mu}, V_{(k), \mu} ) {p_{s_\mu}^{(k)}}(s_\mu), \\
    &\propto \, s_\mu \, \exp\Big\{-\frac{(s_\mu- \ms_{(k), \mu})^2}{2\sigs_{(k), \mu}}\Big\}{p_{s_\mu}^{(k)}}(s_\mu)
\end{align}
so that 
\begin{align}
    {p_{s_\mu}^{(P)}}(s_\mu) &\propto \, s_\mu^{P} \, \exp\Big\{-\frac{(s- \Lambda_{(P), \mu})^2}{2\Sigma_{(P), \mu}}\Big\} \quad  \text{ with } 
    \left\{
    \begin{array}{l}
    \Sigma_{(P), \mu}^{-1} = \sum\limits_{k=1}^P \sigs_{(k), \mu}^{-1} \\
    \Lambda_{(P), \mu} = \Sigma_{(P), \mu}\left(  \sum\limits_{k=1}^P \sigs_{(k), \mu}^{-1} \ms_{(k), \mu} \right)\, ,
    \end{array}
    \right.
\end{align}
yielding a posterior on the calibration variable at step $P$ with an identical form to the posterior of the offline algorithm (albeit with parameters computed differently). Therefore $\sh_\mu$ and $\Cs_\mu$ can still be computed analytically following \eqref{eq:chap6-gain-sh} and \eqref{eq:chap6-gain-Cs}. We provide a pseudo-code for the online cal-AMP in \citealg~\ref{alg:chap6-online-cal-amp} and a pseudo-code for online SE in \citealg~\ref{alg:chap6-online-cal-se}. 

\begin{figure}[t]
    \centering
    \includegraphics[width=\textwidth]{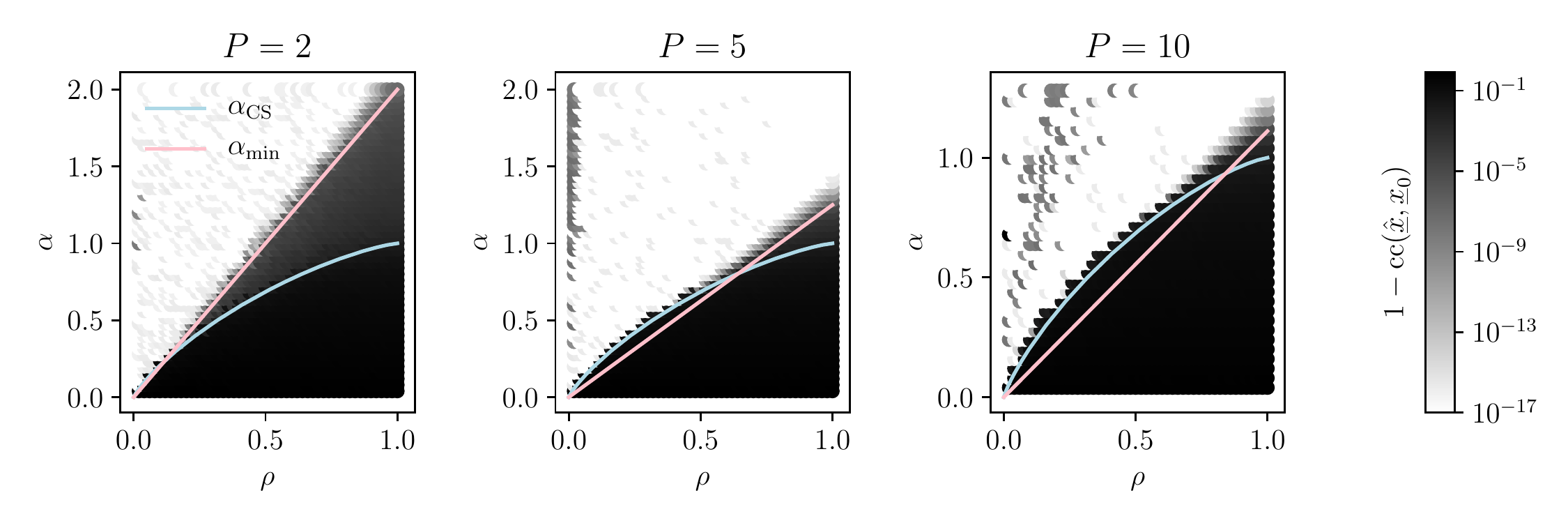}
    \includegraphics[width=\textwidth]{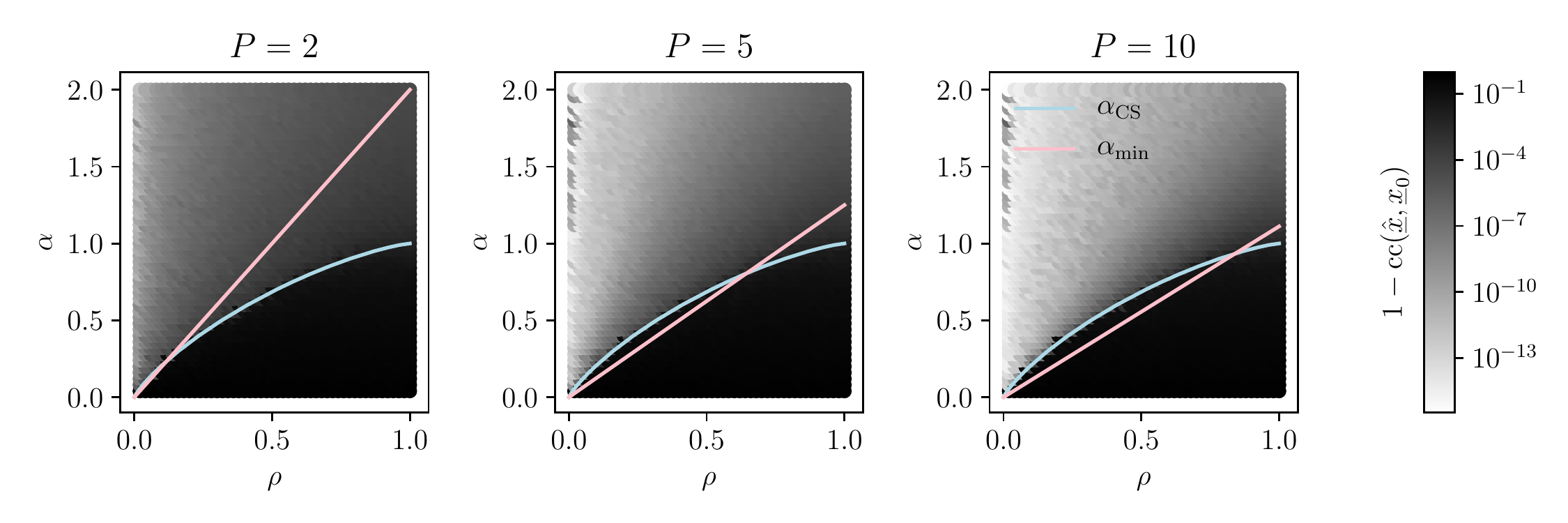}
    \caption{Normalized cross correlation between cal-AMP estimate $\hat{\X}$ and teacher signal $\X_0$ for the gain calibration problem with calibration variables uniformly distributed in $[0.95,1.05]$ and $N=10^3$. Diagrams are plotted as a function of the measurement rate $\alpha = M/N$ and the sparsity level $\rho$ for the offline \textbf{(top row)} and online \textbf{(bottom row)} algorithms, for increasing number of available samples $P$. The blue line $\alpha_{\rm CS}$ is the phase transition threshold for a perfectly calibrated channel. The pink line $\alpha_{\rm min}$ marks the strict lower bound on the number of measurements necessary for reconstruction. The online cal-AMP requires more samples than the offline version to achieve comparable errors, nevertheless above the transition relatively low errors are already reached at $P=10$.
    We note that the algorithms are sometimes unstable at low $\rho$, leading to unexpectedly high MSEs (top left corner of top right diagram).   \label{fig:chap6-calamp-mul-diag-x}}
\end{figure}

\begin{figure}[t]
    \centering
    {\includegraphics[width=\textwidth]{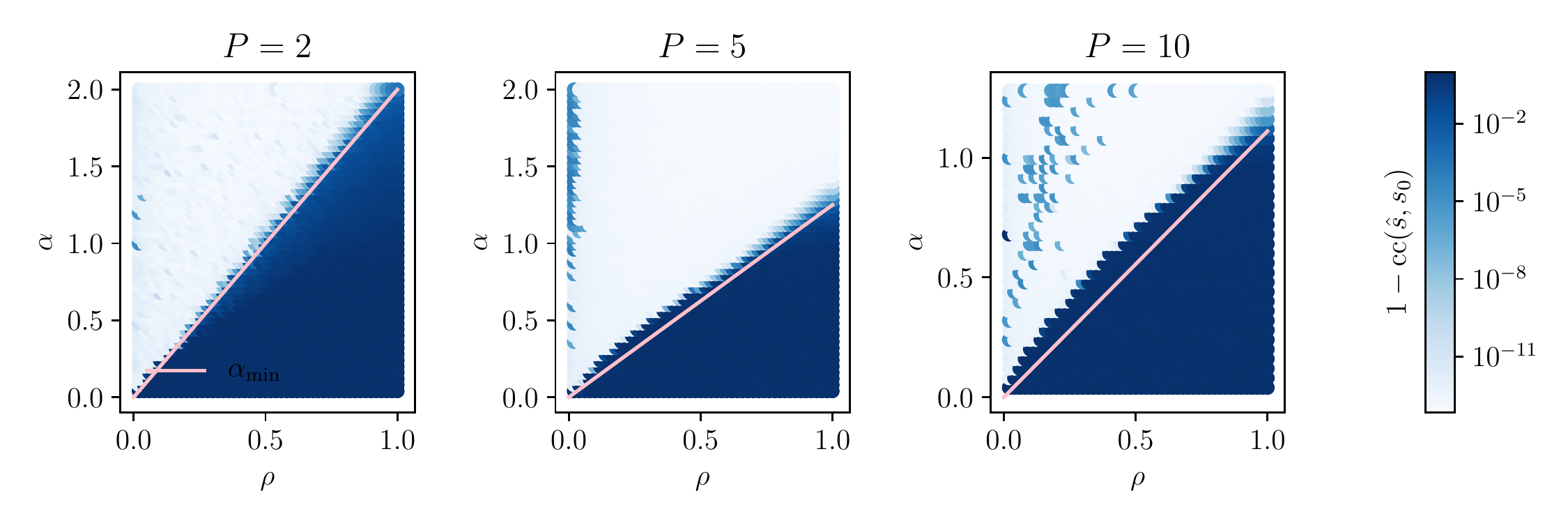}
    \label{fig:chap6-calamp-mul-diag-offline-s}}
    {\includegraphics[width=\textwidth]{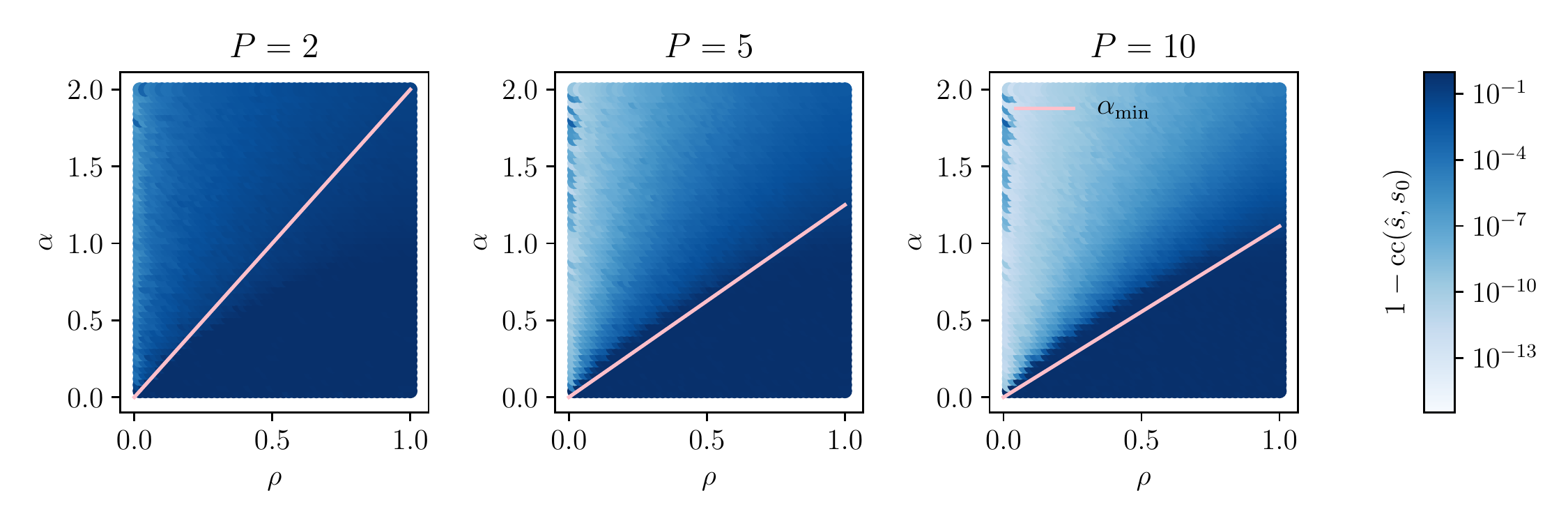}
    \label{fig:chap6-calamp-mul-diag-online-s}}
    \caption{Normalized cross correlation between cal-AMP estimate $\hat{\s}$ and teacher signal $\s_0$ for the gain calibration problem with calibration variables uniformly distributed in $[0.95,1.05]$ and $N=10^3$. Diagrams are plotted as a function of the measurement rate $\alpha = M/N$ and the sparsity level $\rho$ for the offline \textbf{(top row)} and online \textbf{(bottom row)} algorithms, for increasing number of available samples $P$. The pink line $\alpha_{\rm min}$ marks the strict lower bound on the number of measurements necessary for reconstruction. Results are similar to the diagrams in terms of errors on the signal $\hat{X}$ of \citefig~\ref{fig:chap6-calamp-mul-diag-x}.  \label{fig:chap6-calamp-mul-diag-s}}
\end{figure}

On the bottom rows of \citefig~\ref{fig:chap6-calamp-mul-diag-x} and \ref{fig:chap6-calamp-mul-diag-s} we plot phase diagrams obtained with online cal-AMP. Compared to the offline diagrams, we find that the reconstruction requires more samples to achieve comparable levels of accuracy.  On \citefig~\ref{fig:chap6-calamp-mul-dyn-online} and \ref{fig:chap6-calamp-mul-comp-online} we check the consistency of the cal-AMP and SE fixed points in terms of the MSE of cal-AMP in reconstructing $\X_0$.

\vspace{3cm}

\pagebreak

\begin{algorithm*}[ht]
\caption{Online Gain Calibration Approximate Message Passing\label{alg:chap6-online-cal-amp}}   
\begin{algorithmic}
    \State {\bfseries Input:} matrix $\Y \in \R^{M \times P}$ and matrix $\W \in \R^{M \times N}$:
    \State \emph{Initialize:} $\Lambda_{0,\mu}=0$, $\Sigma_{0, \mu}=0$,
    \For{$k = 1, \cdots, P:$}
        \State \emph{Initialize}: $\hat{x}_{i,k}$, $C^x_{i,k} \quad \forall\ i$ and $\gouts_{\mu, k}$, $\dgouts_{\mu, k} \quad \forall\ \mu$
        \Repeat   
        \Statex \hspace{0.35cm} 1) Estimate mean and variance of $\z\kk$ given current $\hat{\x}\kk$
        \vspace{-0.3cm}
        \begin{align}
            V_{\mu, k}^{(t)} &= \sum\limits_{i=1}^N W_{\mu i}^2 {C^x_{i,k}}^{(t)} \label{alg:chap6-online-cal-amp-V} \\
            \omega_{\mu, k}^{(t)} &= \sum\limits_{i = 1}^N W_{\mu i} \hat{x}^{(t)}_{i,k} - \sum\limits_{i = 1}^N W_{\mu i}^2 {C^x}^{(t)}_{i,k} \gouts_{\mu, k}^{(t-1)} \label{alg:chap6-online-cal-amp-om}
        \end{align} 
        \vspace{-0.3cm}
        \Statex \hspace{0.35cm} 2) Exploit current $\y\kk$ and inherited $\Lambda_{k-1, \mu},\Sigma_{k-1, \mu}$

        \Statex \hspace{0.6cm} 2.1) Update recursion parameters given $\y\kk$
        \vspace{-0.3cm}
        \begin{align}
            {\sigs_{\mu, k}^{(t)}}^{-1} &= y_{\mu, k}^2 / (V_{\mu, k}^{(t)} + \Delta)\\
            \ms_{\mu, k}^{(t)} &= \sigs_{\mu, k} y_{\mu, k}\omega_{\mu, k}^{(t)} / (V_{\mu, k}^{(t)} + \Delta)
        \end{align}
        \vspace{-0.5cm}
        \begin{align}
            \Sigma_{k, \mu}^{(t)} &= \Sigma_{k-1, \mu} + \sigs_{\mu, k}^{(t)} \\
            \Lambda_{k, \mu}^{(t)} &=\Sigma_{k, \mu} \left( \Sigma_{k-1, \mu} \Lambda_{k-1, \mu} + \sigs_{\mu, k}^{(t)}\ms_{\mu, k}^{(t)} \right)
        \end{align}

        \Statex \hspace{0.6cm}  2.2) Update estimates $\sh$ and $\Cs$
        \begin{align}
            \sh^{(t)}_\mu & = \frac{\mathcal{I}(k+1, \Lambda_{k, \mu}, \Sigma_{k, \mu}, a, b)}{\mathcal{I}(k, \Lambda_{k, \mu}, \Sigma_{k, \mu}, a, b)} \\
            {\Cs}^{(t)}_\mu & = \frac{\mathcal{I}(k+2, \Lambda_{k, \mu}, \Sigma_{k, \mu}, a, b)}{\mathcal{I}(k, \Lambda_{k, \mu}, \Sigma_{k, \mu}, a, b)}
        \end{align}

        \Statex \hspace{0.6cm}  2.3) Update $\gout$ and $\dgout$
        \vspace{-0.3cm}
        \begin{align}
            \dgouts^{(t)}_{\mu, k} &=  {\Cs}^{(t)}_\mu  y_{\mu,k}^2 / (V_{\mu, k}^{(t)} + \Delta )^{2} - (V_{\mu, k}^{(t)} + \Delta )^{-1}\label{alg:chap6-online-cal-amp-dg} \\
            \gouts^{(t)}_{\mu, k} &= (\sh_\mu^{(t)}y_{\mu,k} - \omega_{\mu,k}^{(t)}) / (V_{\mu, k}^{(t)} + \Delta ),  \label{alg:chap6-online-cal-amp-g}
        \end{align}
        \vspace{-0.5cm}
        \Statex \hspace{0.35cm} 3) Estimate mean and variance of $\x$ given current optimal $\z$
        \vspace{-0.3cm}
        \begin{align}
            \sigma_{i,k}^{(t)} &= \left(- \sum\limits_{\mu=1}^{M}W_{\mu i}^2\dgouts_{\mu, k}^{(t)}\right)^{-1} \label{alg:chap6-online-cal-amp-sig} \\
            \lambda_{i,k}^{(t)} &=  \hat{x}^{(t)}_{i,k} + \sigma_{i,k}^{(t)}\left( \sum\limits_{\mu=1}^{M}W_{\mu i}\gouts_{\mu, k}^{(t)}\right) \label{alg:chap6-online-cal-amp-lbd}
        \end{align}
        \vspace{-0.3cm}
        \Statex  \hspace{0.35cm} 4) Estimate of mean and variance of  $\x$ augmented of the information about the prior
        \vspace{-0.3cm}
        \begin{align}
            {C^x_{i,k}}^{(t+1)} &= f^x_2(\lambda^{(t)}_{i,k}, \sigma_{i,k}^{(t)}) \label{alg:chap6-online-cal-amp-cx} \\
            \hat{x}^{(t+1)}_{i,k} &= f^x_1(\lambda^{(t)}_{i,k}, \sigma_{i,k}^{(t)})  \label{alg:chap6-online-cal-amp-xh}
        \end{align}
        \Statex \hspace{0.35cm} $t = t+1$
        \vspace{0.15cm}
        \vspace{0.05cm}
        \Until{convergence} 
    \EndFor
    \State {\bfseries Output:} Estimates and variances $\hat{\x}_i$, $\vect{C}^x_i$, $\sh_\mu$, $\Cs_\mu$
\end{algorithmic}
\end{algorithm*}





\section{Conclusion}

In this paper, we presented a theoretical characterization of the cal-AMP algorithm for blind calibration. The derivation of the corresponding State Evolution was presented from the message passing equations of a vectorial version of the GAMP algorithm. Furthermore, we extended the algorithm and its analysis to the online scenario, which allows to adapt the calibration upon receiving new observations. The validity of the newly proposed algorithms and analysis were demonstrated through numerical experiments on the case of gain calibration.
In the offline setting, cal-AMP achieves near-perfect reconstruction close to the theoretical lower bound. In the online setting, cal-AMP is also able to calibrate and reconstruct from a small number of samples.

For practical applications, the extension to the online case is of particular interest. It enables continuous improvement of hardware sensing channel, without requiring to keep past observations in memory. A natural extension will also be to consider the G-VAMP \cite{Rangan2016, Schniter2016} version of cal-AMP, which will enable a generalization of the algorithm to measurement matrices with non-i.i.d. entries. 
From the theoretical point of view, we note that the introduced SE analysis in the non-Bayes optimal setting should make it possible to characterize other reconstruction algorithms, as algorithms proposed in \cite{Gribonval2012,Shen2013}. 
Also, it will be interesting to confirm whether cal-AMP can also succeed in a multi-layer setting, likewise the recent extension of GAMP to ML-AMP \cite{Manoel2017b}. This generalization will open the way to message passing algorithms solutions in more complex calibration problems.

\paragraph{Acknowledgements}
We wish to thank Christophe Sch\"ulke for his dedication to the calibration
problem. We would like to thank the Kavli Institute For Theoretical Physics for its hospitality for an extended stay, during which parts of this work were conceived and carried out. This work is supported by the French Agence Nationale de la Recherche under grant ANR17-CE23-0023-01 PAIL and the European Union’s Horizon 2020 Research and Innovation Program 714608-SMiLe. This material is based upon work supported by Google Cloud. We gratefully acknowledge the support of NVIDIA Corporation with the donation of the Titan Xp GPU used for this research. Additional funding is acknowledged by MG from ‘Chaire de recherche sur les modèles et sciences des données’, Fondation CFM pour la Recherche-ENS.

\bibliographystyle{unsrt}
\bibliography{paper-cal-amp.bib}

\end{document}